\documentclass[twocolumn]{aastex631}

\usepackage{xcolor}
\usepackage{refcount}
\usepackage{booktabs}
\usepackage{amsmath}
\usepackage{ulem}

\shorttitle{Paloma}
\shortauthors{Dutta and Rana}
\graphicspath{{./}{figures/}}

\begin{document}

\title{Study of Complex Absorption and Reflection in a Unique Intermediate Polar Paloma}

\correspondingauthor{Anirban Dutta}
\email{anirband@rri.res.in}

\author[0000-0001-9526-7872]{Anirban Dutta}
\affiliation{Astronomy and Astrophysics Group, Raman Research Institute, C. V. Raman Avenue, Sadashivanagar, Bengaluru, KA-560080, India}

\author[0000-0003-1703-8796]{Vikram Rana}
\affiliation{Astronomy and Astrophysics Group, Raman Research Institute, C. V. Raman Avenue, Sadashivanagar, Bengaluru, KA-560080, India}








\begin{abstract}

We present the broadband  (0.3-40.0 keV) X-ray analysis of a unique intermediate polar Paloma using simultaneous data from XMM-Newton and NuSTAR observatories. The X-ray power spectra show strong modulations over orbital period compared to spin period. The orbit folded lightcurves show single broad hump like structure with strong dips for soft to medium X-rays (0.3-10.0 keV). The energy dependent dips at $\phi \sim 0.16$ and $0.5$ arise due to complex intrinsic absorber, strong enough to have effect well around 15 keV. The absorber could potentially be contributed from accretion curtain/ accretion stream and absorbing material produced by stream-disc/stream-magnetosphere interactions. We notice significant variation of the absorber with orbital phase, with maximum absorption during orbital phase 0.1-0.22. The absorber requires more than one partial covering absorber component, specifying the necessity to use distribution of column densities for spectral modelling of the source. Isobaric cooling flow component is utilized to model the emission from the multi-temperature post-shock region, giving shock temperature of $31.7_{-3.5}^{+3.3}$keV, which corresponds to white dwarf mass of $0.74_{-0.05}^{+0.04}\;M_{\odot}$. We have used both the neutral absorber and the warm absorber models, which statistically give similarly good fit, but with different physical implications. Among the Fe K$_{\alpha}$ line complex, the neutral line is the weakest. We probed the Compton reflection, and found minimal statistical contribution in the spectral fitting, suggesting presence of weak reflection in Paloma.

\end{abstract}


\section{Introduction}

Intermediate polars (IPs) are a type of magnetic cataclysmic variables (mCVs) in which the white dwarf (WD) magnetic field is relatively weaker than the polars. Unlike the polars, the magnetic locking between the WD, and the companion star is absent in IPs, giving rise to lack of synchronicity between the spin period ($P_\omega$) and the orbital period ($P_\Omega$). In most IPs, the spin period is much shorter than the orbital period ($P_\omega \lesssim 0.1 \times P_\Omega$ ).

In IPs, a weaker magnetic field may allow formation of partial accretion disk before the magnetic field disrupts the disk. During the final stretch of the accretion, the accreting material falls on the WD following the magnetic field lines of the WD. The infalling material with supersonic velocity creates strong shock over the WD surface. Subsequently, the material cools down in the post shock region (PSR) by radiating bremsstrahlung emission in X-rays and cyclotron emissions in optical wavelengths, and finally settles down on the WD surface. The X-rays emitted from the PSR interacts with the intrinsic absorber in the source like accretion stream or accretion curtain as well as any intervening medium like the interstellar medium (ISM) while reaching the observer. A part of the X-rays can get Compton reflected by the WD surface and produce excess (hump-like feature) in 10-30 keV range of the spectrum. Also, a part of the emitted X-rays will undergo photoelectric absorption, giving rise to fluorescence emissions, most notably iron fluorescence line at 6.4 keV line. It is believed that the strength of the iron fluorescence line is correlated to the Compton reflection as both the physical processes can take place in similar region of WD surface.

The source Paloma (RX J0524+42, 1RXS J052430.2+424449) is a special IP in which the asynchronism is much less ($\sim 14\%$)  compared to other IPs (typically $\sim 90\%$). \cite{Schwarz_2007} has studied the source using optical photometry on multiple nights, extended over a duration of several years and ROSAT X-ray observation. They detected three strong periodic signatures in the optical lightcurve and commented about two possible values of the spin period. Recently, \cite{Littlefield_2022} specified one of the two values of \citet{Schwarz_2007} for the spin period of the WD using nearly a month long TESS observation. This source was studied by \cite{Joshi_2016} in the soft to medium X-rays (0.3-10 keV) using XMM-Newton. Their spectral modelling hinted at the presence of strong reflection amplitude. Paloma is yet to receive its proper classification among mCVs. Here we should note that rate of change of spin period ($\dot{P}$) is not yet determined for Paloma. Given the asynchronosity which is much lower than conventional IPs and slightly higher than traditional asynchronous polars (APs) ($\sim1-2\%$ eg. V1432 Aql, CD Ind, BY Cam, V1500 CYg) for which $\dot{P}$ is measured, it is hard to distinctly classify Paloma as IP or AP. \citet{Schwarz_2007, Joshi_2016} described Paloma as IP whereas \citet{Littlefield_2022} considered it as AP.

Using the broadband X-ray data, simultaneously obtained from XMM-Newton and NuSTAR, we tried to give an overall description of the spectral properties of the system, along with the temporal behavior of the system extending all the way up to 40 keV. The strong reflection amplitude should manifest a reflection hump in the 10-30 keV which we could not unambiguously detect in the NuSTAR spectra. We argued for an alternative picture to the strong reflection hypothesis, using a complicated absorber model for the source. We also probed for X-ray periodicity in the power spectra. In the following section we describe the reduction of the obtained data. In Sec. \ref{Sec:analysisresults}, we present our results based on the timing and the spectral analysis of the broadband X-ray data. In the next section (Sec. \ref{Sec:discussion}) we discuss our results and we summarise the results in the concluding section (Sec. \ref{Sec:conclusion}).

\section{Observation and Data reduction}
\label{Sec:obsdatareduction}
Paloma was observed simultaneously by the XMM-Newton \citep{Jansen_2001} and NuSTAR \citep{Harrison_2013} observatories. XMM-Newton can observe the source in 0.3-10.0 keV band with good spectral resolution whereas NuSTAR is capable of imaging the source in 3.0-78.0 keV band with high sensitivity in the hard X-rays. Thereby simultaneous observation by these two observatories provides us broadband data in 0.3-78.0 keV. This enables us to probe the properties of the source in the soft X-rays like constraining the absorption parameters as well as detecting the shock temperature and Compton reflection in the hard X-rays. 

NuSTAR observed (Obs ID 30601019002) the source on UTC 2021-03-02 01:10:08. The focal plane modules (FPMA and FPMB) recorded the source for an exposure time of $\sim 47$ks. We have selected a source region using a circle of radius 40 arcsec and a background region of radius 80 arcsec in the same detector as that of the source. We used NUPIPELINE to extract the cleaned events files with default screening criteria and then NUPRODUCTS to obtain the final science products like lightcurves and the spectra along with detector response files. We performed barycenter correction while deriving the science products. The timebin size of the extracted lightcurves are chosen to be 1s. We have grouped the spectra for a minimum counts of 25 per bin to use $\chi^2$ statistics to test goodness-of-fit.

The XMM-Newton observation (Obs ID 0870800201) was taken on UTC 2021-03-02 01:06:28 for a duration of $\sim 33$ks. The observation was carried in full window mode using medium filter for both of the PN \citep{Struder_2001} and MOS \cite{Turner_2001} cameras of European Photon Imaging Camera (EPIC) instruments. We have selected a circular source region of radius 25 arcsec and a circular background region of radius 50 arcsec in the same detector as that of the source. We used XMMSAS v19.1.0 \citep{Gabriel_2004} for data reduction and extraction of the lightcurves and the spectra following the SAS analysis thread \footnote{\url{https://www.cosmos.esa.int/web/xmm-newton/sas-threads}}. The calibration files are used from the current calibration file (CCF) repository \footnote{\url{https://www.cosmos.esa.int/web/xmm-newton/current-calibration-files}} of XMM-Newton, latest during the analysis.  The XMM data is contaminated with background flaring in the initial part of the observation, so we discarded first few ks of data which neatly eliminate the flaring part and leaves us with nearly $\sim 29$ks of continuous data from all the EPIC cameras. Using SAS tool EPATPLOT, we have also checked for pile up in the data, and did not find any significant contribution of it. The products have been barycenter corrected. We chose the 1s binning for the extracted lightcurves. The SAS tool SPECGROUP is used to group the data with minimum counts per bin at least 25 for using the $\chi^2$ statistics. The oversample parameter is set at 3 so that minimum group width is not less than 1/3rd of the FWHM of the instrument resolution at that energy.

We have also extracted the spectrum from reflection grating spectrometer (RGS) on board XMM-NEWTON. We have used RGSPROC tool and eliminated the background flaring to obtain the grating spectra. The spectra are also grouped to 25 minimum counts per bin to use $\chi^2$ statistics.

\section{Data Analysis and Results}
\label{Sec:analysisresults}

\subsection{Timing Analysis}

\begin{figure}
   \centering
	\includegraphics[width=\columnwidth, trim = {0.5cm 0cm 0.1cm 0cm}, clip] {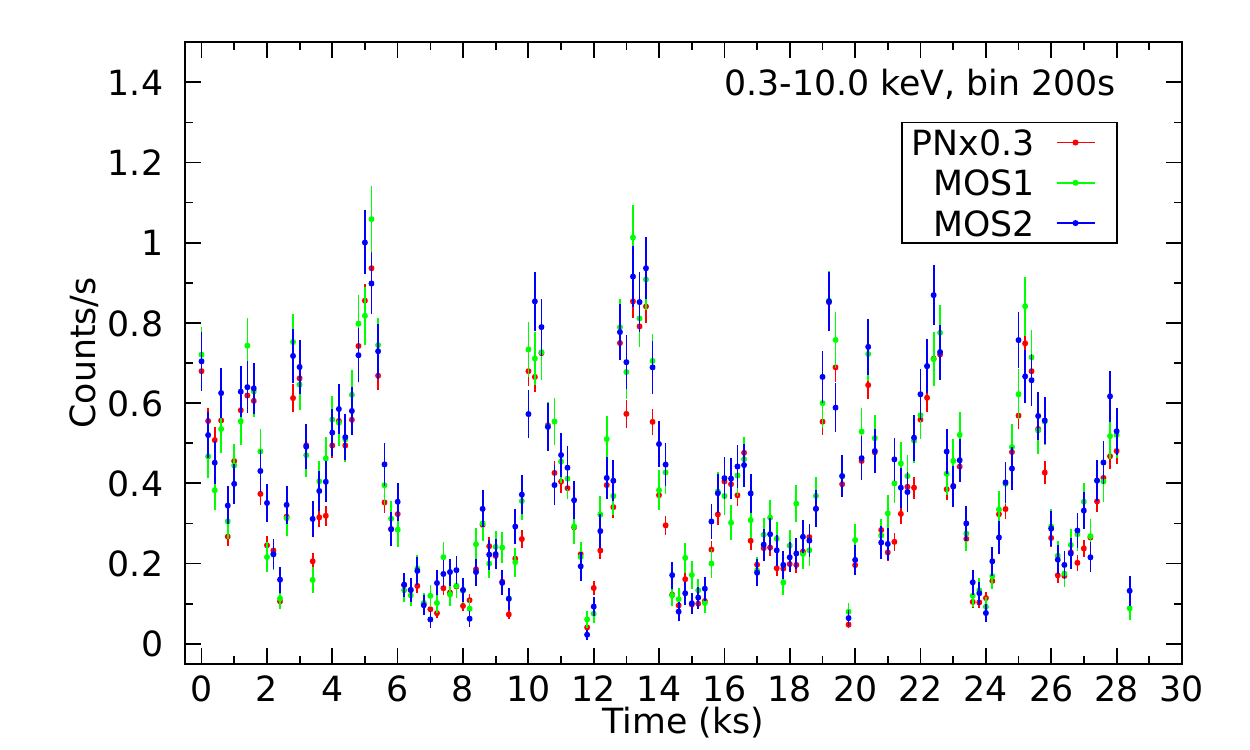}
	\includegraphics[width=\columnwidth, trim = {0.5cm 0cm 0.1cm 0cm}, clip]{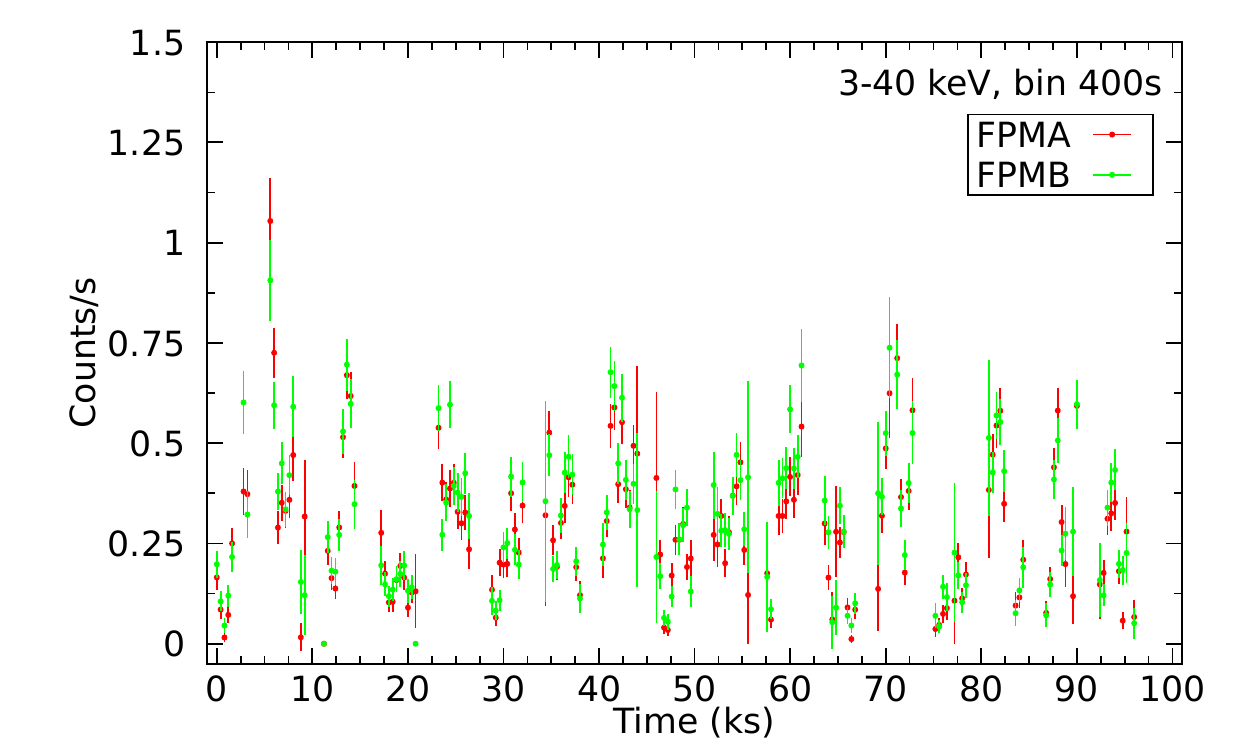}
	\includegraphics[width=\columnwidth, trim = {0.5cm 0.1cm 0.1cm 0.1cm}, clip]{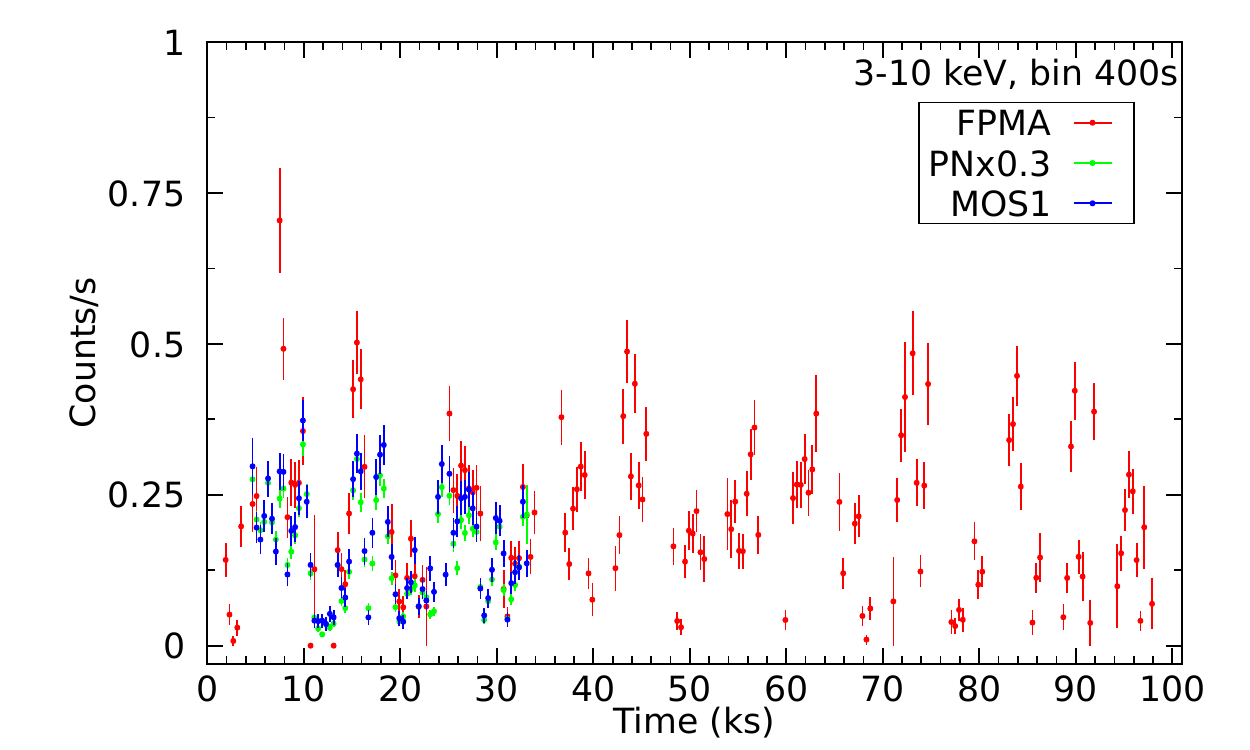}      
	\caption{Background subtracted lightcuves from XMM-Newton (top), and NuSTAR (middle) and overlapping 3-10 keV lightcurves from both the observatories for comparison (bottom).}
         \label{Fig:cleanlightcurves}
   \end{figure}

We have the continuous data from XMM-Newton for $\sim 29$ks and a total duration of $\sim96$ ks data from NuSTAR including the gaps due to earth occultation and South Atlantic Anomaly (SAA). The duration of XMM-Newton and NuSTAR observation give us the opportunity to study the timing properties of the source for $\sim3$  cycles and  $10$ cycles respectively. We have shown the background subtracted lightcurves from XMM-Newton and NuSTAR in the top and middle panels of Fig. \ref{Fig:cleanlightcurves}. In the bottom panel of Fig \ref{Fig:cleanlightcurves}, the lightcurves in the 3-10 keV band from both the observatories are overplotted. In this section, we have calculated the power spectra for the source, and subsequently folded the lightcurves based on the orbital period of the system. We have presented the results from NuSTAR-FPMA and XMM Newton EPIC-PN in the following subsections; as the results from FPMB for NuSTAR and MOS1 \& MOS2 for XMM Newton EPIC are also similar.

\subsubsection{Power Spectra}

We have implemented two methods to search for the periods - Lomb-Scargle method \citep{Lomb_1976, Scargle_1982} and the CLEAN algorithm \citep{Roberts_1987}. The first one is widely used to find the frequencies present in a unevenly sampled time series. The later one uses the algorithm to deconvolve the effect of window function from the discrete Fourier transform (DFT) spectra and thereby minimise the effect of spectral leakage. This method has been used efficiently for other IP sources with unevenly sampled data \citep{Norton_1992, Norton_1997}.

\begin{figure}
   \centering
	\includegraphics[width=\columnwidth, trim = {0.1cm 0.1cm 0.1cm 0.1cm}, clip]{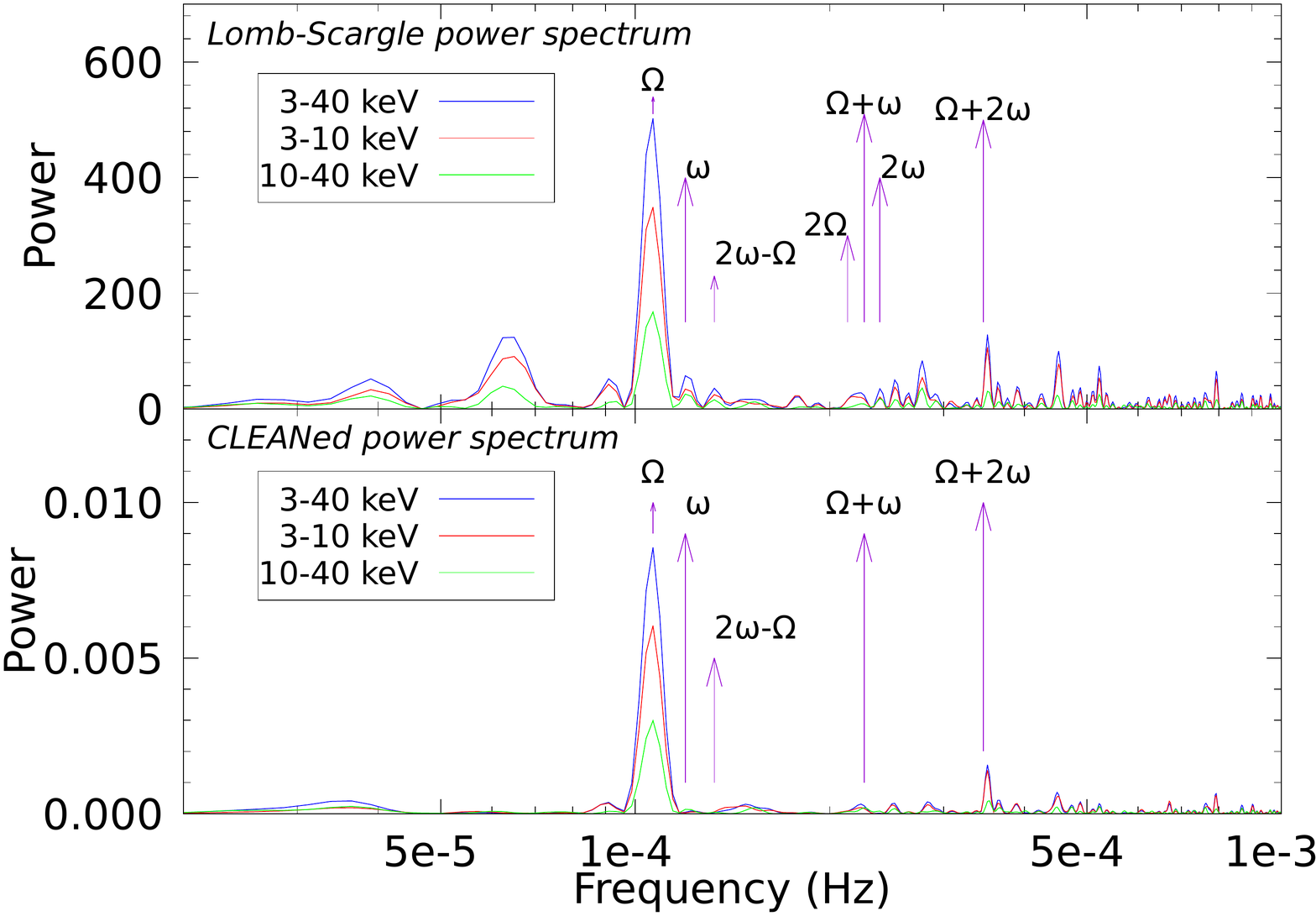}
		\caption{Power spectra obtained from NuSTAR-FPMA using Lomb-scargle algorithm (\textcolor{blue}{top figure}), and CLEAN algorithm (\textcolor{blue}{bottom figure}). We have identified the two fundamental peaks, $\Omega$ and $\omega$, along with the location of few sideband and harmonics.}
         \label{Fig:fpmapowerspec}
   \end{figure}

In the NuSTAR-FPMA power spectra using Lomb-Scargle method (top panel of Fig. \ref{Fig:fpmapowerspec}), we can clearly identify the peak at orbital frequency $\Omega$, corresponding to a period value $9384\pm229$s, present in all the energy bands (3-40 keV, 3-10 keV, and 10-40 keV). We notice that the peaks are stronger in 3-10 keV band compared to 10-40 keV band. This period is also identified by \cite{Schwarz_2007} at $\sim9430.35$s in their optical photometric campaign and \citet{Thorstensen_2017} at $\sim 9434.88$s using radial velocity spectroscopy. We see the spin frequency peak $\omega$, corresponding to a period value $8365\pm181$s, immediately next to the orbital frequency in the Lomb-Scargle power spectra. This value agrees with one of the spin period candidate ($\sim 8175.37$s) by \cite{Schwarz_2007} , and spin period ($\sim 8229s$) detected by \cite{Littlefield_2022}. We notice another weak peak at $2\omega-\Omega$, with a period value of $7545\pm148$s. In the FPMA power spectra, these two peaks ($\omega, 2\omega-\Omega$) appear with a much lower power compared to the $\Omega$.  \cite{Joshi_2016} obtained two low power peaks for $\omega$ and $2\omega-\Omega$ adjacent to the strongest peak at orbital period in their power spectra, obtained using previous XMM-Newton observation, but with smaller period values (7800s and 6660s respectively) than ours. In Fig. \ref{Fig:fpmapowerspec} we have also marked the expected positions of several harmonics and sideband frequencies. We notice a relatively strong peak ($2850\pm21$s) near $\Omega+2\omega$ (expected at $2893\pm48$s). We also observe peaks at $4474\pm52$s and $4183\pm46$s which respectively agree with $\Omega+\omega$ (expected at $4423\pm72$) and $2\omega$ (expected at $4182\pm90$s) within error-bar. It is to be pointed here that, the resolution of our power spectra which depends upon the total observation duration, data gaps etc, create hindrance to resolve all the harmonics and sidebands precisely. The another possible candidate of spin period at $\sim8758.12$s according to \cite{Schwarz_2007}, which is very close to peak $\Omega$, could not be confirmed due to the limitation of the resolution of the power spectra. We also notice that the peaks in the FPMA power spectra are stronger in 3-10 keV band compared to 10-40 keV band.

In the bottom panel of Fig. \ref{Fig:fpmapowerspec}, we have plotted the CLEANed power spectra. Before performing the CLEAN operation, we have detrended the lightcurve for the presence of dc bias, if any, by subtracting a mean value and dividing by the standard deviation. The CLEANed algorithm requires two parameters - number of iterations and loop gain - to deconvolve the window power spectra from the DFT spectra iteratively with the gain fraction in each iteration . We chose gain 0.1, typically used \citep{Norton_1992, Rana_2004} for this algorithm. Iteration number was set at 1000. We find that, the strong $\Omega$ peak and the $\Omega+2\omega$ peak is present, however, other weak peaks at $\omega, 2\omega-\Omega$ and $2\omega$  which were in seen in Lomb-Scargle power spectra are almost diminished in the CLEANed power spectra.

\begin{figure}
   \centering
	\includegraphics[width=\columnwidth, trim = {0.1cm 0.1cm 0.1cm 0.1cm}, clip]{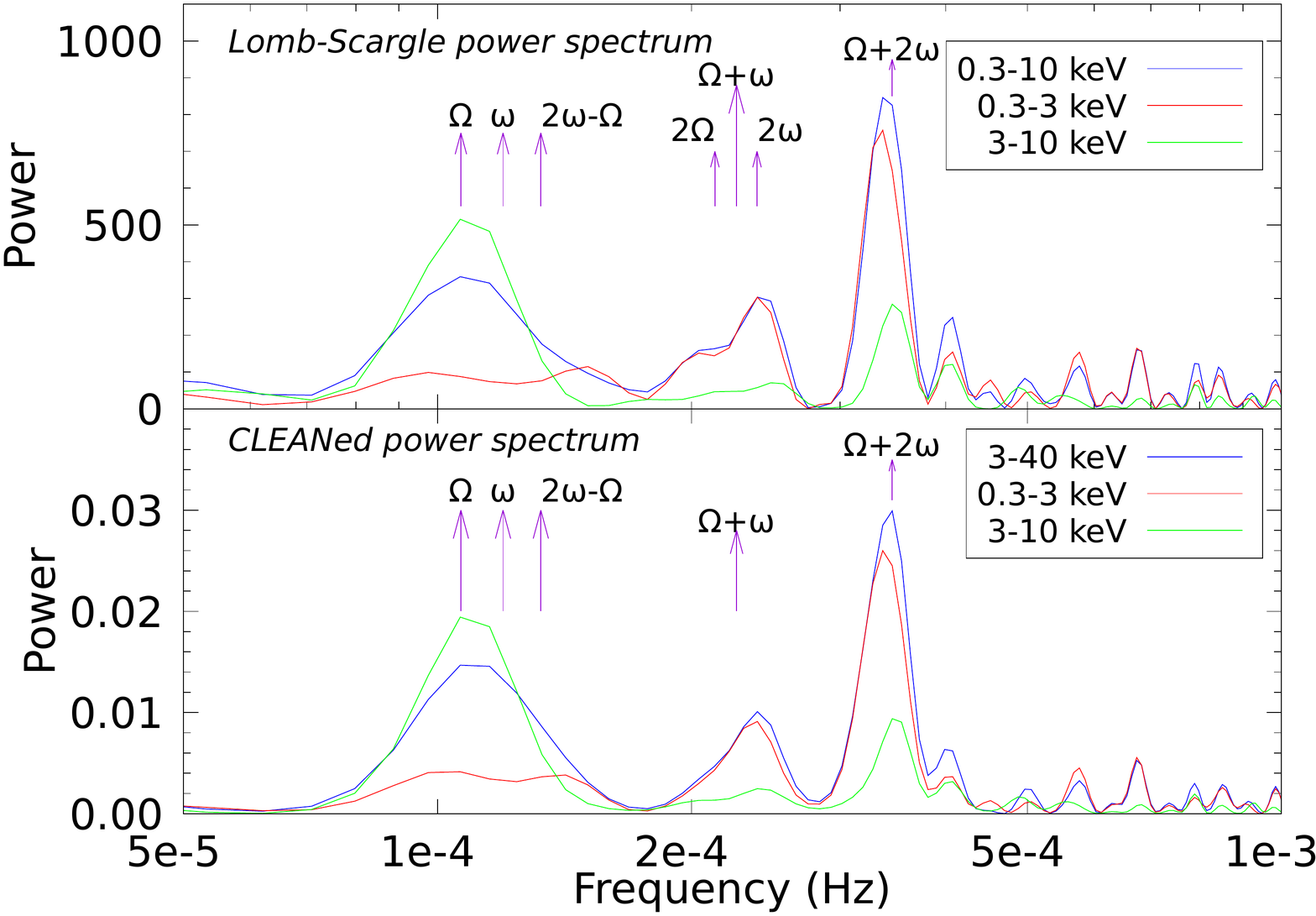}
	\caption{Power spectra obtained from XMM-Newton PN using Lomb-Scargle algorithm (\textcolor{blue}{top figure}), and CLEAN algorithm (\textcolor{blue}{bottom figure}). The $\Omega$ peak is identified along with expected (from FPMA power spectra) position of $\omega$ is marked. Few sideband frequencies and harmonics are labelled. }
         \label{Fig:pnpowerspec}
   \end{figure}

The similar exercise is done for the XMM-Newton PN power spectra (Fig. \ref{Fig:pnpowerspec}). Though the PN lightcurves are continuous, the power spectral resolution is poor due to shorter duration of the observation. We have found that the peak corresponding to $\Omega$ is evidently present, with a period value $9400\pm761s$, agreeing with that of the FPMA value. We are unable to distinguish neither of $\omega$ and $2\omega-\Omega$ in the PN power spectra, because of the broad peak of $\Omega$. It is interesting to note that the peak strength of $\Omega$ is much lower in 0.3-3.0 keV band than the 3-10 keV band. However, for all other peaks in the power spectra, peak strength decreases in higher energy band. By our previous definition of orbital and spin periods from FPMA, we have marked their expected positions in the PN power spectra, along with harmonics and sideband frequencies. We notice that there is significantly strong power near the position marked by $\Omega+2\omega$, as we found from FPMA power spectra.

\subsubsection{Folded Lightcurves}
\label{Sect:Folded Lightcurves}

\begin{figure*}
    \centering
	\includegraphics[width=\columnwidth, trim = {0.7cm 1.7cm 4cm 2.6cm}, clip]{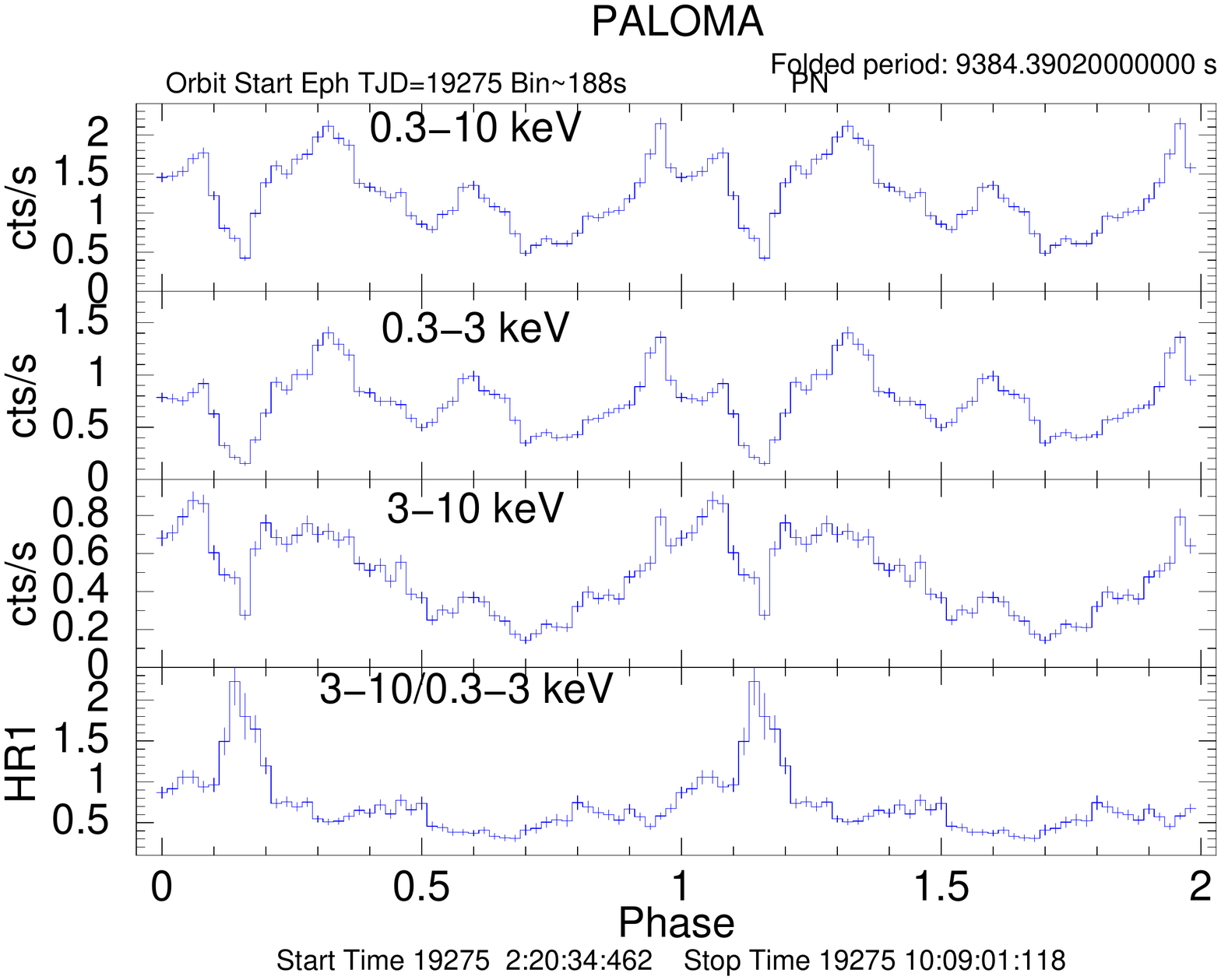}
	\includegraphics[width=\columnwidth, trim = {1.2cm 1.7cm 3.5cm 2.6cm}, clip]{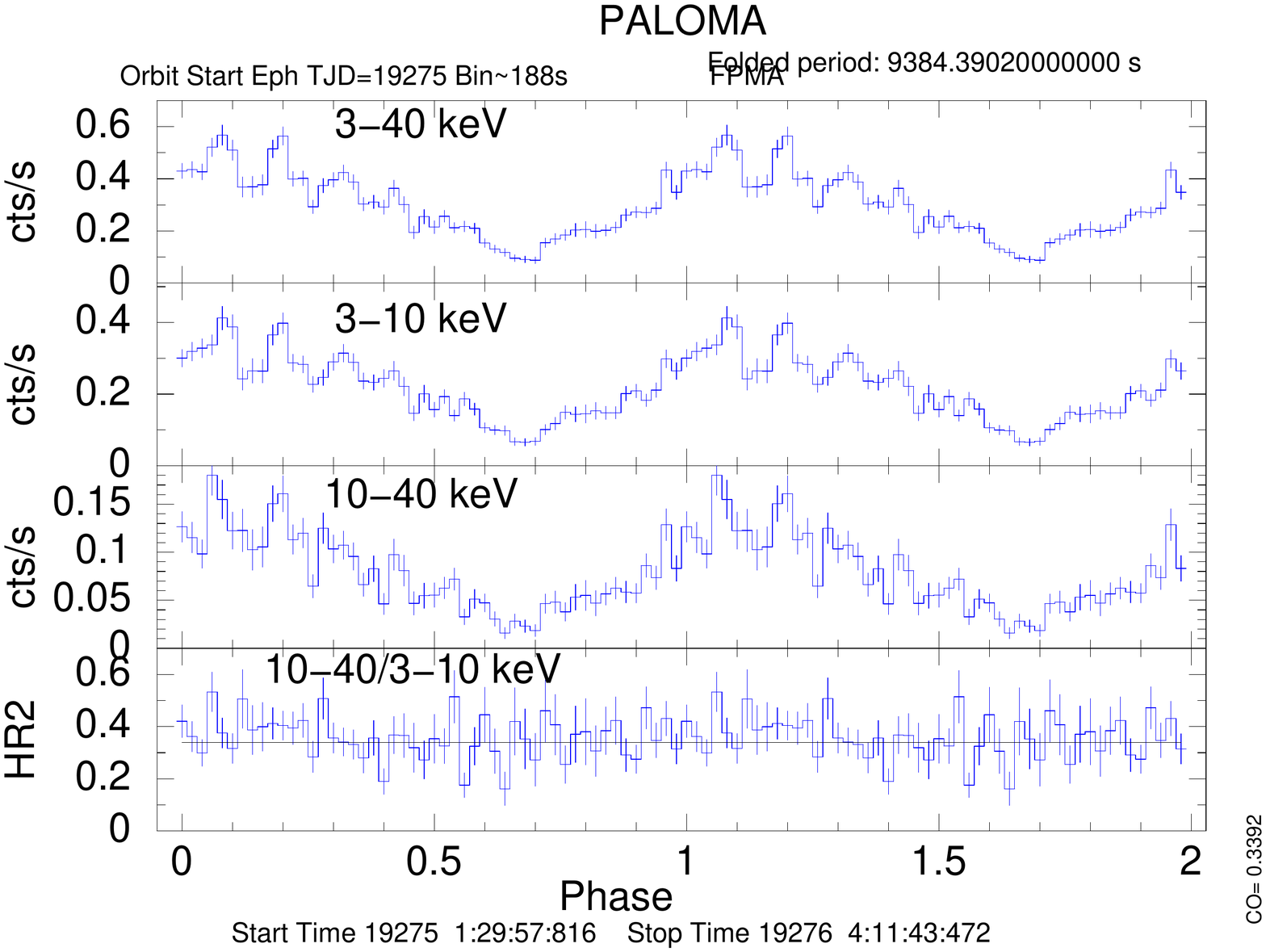}

	\caption{Folded lightcurve, based on period 9384s, from different energy bands, using PN (on the \textcolor{blue}{Left}) and FPMA (on the \textcolor{blue}{Right}) are shown. The bin size is 188s. In the bottom panels, hardness ratio-1 (Left plot) and hardness ratio-2 (right plot) are shown. A \texttt{constant} model is fitted to HR2, represented by a black line.}
         \label{Fig:foldedlightcurve}
   \end{figure*}

Based on our measured orbital period ($9384\pm229$s), we have folded the PN and FPMA lightcurves to study the rotational properties of the source. The folding is performed based on the ephemeris TJD=19275, chosen arbitrarily close to start time of NuSTAR observation. The folded lightcurves in different energy bands and the hardness ratios (HR) are shown in Fig. \ref{Fig:foldedlightcurve}. In the soft X-rays, 0.3-3.0 keV band obtained from PN (left plot of Fig. \ref{Fig:foldedlightcurve}), there is strong modulation in the lightcurve, however the entire lightcurve can be described by one broad hump ($\phi=0.7-1.7$), with several dips in between. There are three prominent minima, appearing at $\phi\sim0.16, 0.50, 0.70$, with the first one being the strongest. In the 3.0-10.0 keV band, the similar broad hump like feature is visible, with the minima at $\phi\sim 0.70$ being the strongest one. A direct comparison of PN lightcurve with the FPMA lightcurve (right plot of Fig. \ref{Fig:foldedlightcurve}) in the same energy band 3-10 keV, shows similar nature of the hump and the presence of a overall minima at $\phi\sim 0.70$. The pulse fraction (PF) of modulation in the 3-10 keV energy band is, $72\pm4\%$ and $72\pm6\%$ respectively from PN and FPMA, using the definition PF $=(I_{max}-I_{min})/(I_{max}+I_{min})$ where $I$ denotes the count rate. The hard X-ray in 3-10 keV band of FPMA follows the same pattern of the lightcurve as that of 10-40 keV. 

As we inspect the hardness ratio-1 (HR1), defined as $\text{HR1}=I_{3-10\;\text{keV}}/I_{0.3-3.0\;\text{keV}}$, obtained from PN lightcurve, we notice a strong peak around $\phi\sim0.16$ (bottom panel of left plot of Fig. \ref{Fig:foldedlightcurve}). This corresponds to the dip at the same phase, seen in the 0.3-3.0 keV and 3.0-10.0 keV band of the PN lightcurve. The HR1 can roughly be divided in four intervals, (a) $\phi\sim0.10-0.22$, (b) $\phi\sim0.22-0.50$, (c) $\phi\sim0.50-0.80$, and (d) $\phi\sim0.80-1.10$. In HR1, the phase interval (a) shows the strong peak whereas interval (c) covers the lowest valley region. The (b) and (d) respectively represents the phase intervals before and after the peak region in the HR1. 

The hardness ratio-2 (HR2), defined similarly as that of HR1, but for 10-40 keV and 3-10 keV band of FPMA, exhibit a rather flat profile (bottom panel of right plot of Fig. \ref{Fig:foldedlightcurve}). We have fit a \texttt{constant} model which gives a mean value of 0.34, with an acceptable fit statistic of $\chi^{2}(DOF)=141.5(100)$. 

We didn't attempt detailed analysis based on the period $\omega$, since this period appears very weakly in the FPMA power spectra and could not be resolved in PN power spectra. 

\subsection{Spectral Analysis}

For the spectral analysis, X-ray Spectral Analysis Software package XSPEC v12.12.0 \citep{Arnaud_1996} is used. We have followed the abundance table defined by Wilms and McCray  \citep{Wilms_2000} and photoelectric cross section table set after \cite{Verner_1996}. All the spectral models used for analysis are part of XSPEC package. The errors on the parameter values are quoted within 90\% confidence levels. In this section we have first attempted the spectral analysis with NuSTAR data and XMM-Newton data separately, to have an idea on how the spectra can be modelled in the hard X-ray part and the soft-X ray part, along with detecting the strong line emissions present in the spectra. Finally we have employed the complete spectral model to jointly fit the broadband spectral data obtained from both the observatories. 

\subsubsection{Phenomenological fit to NuSTAR spectra}
\label{sec:nufit}

The FPMA and FPMB data in 3-40 keV range are employed for probing the shock temperature of PSR and the presence of Compton reflection hump. Beyond 40 keV the background dominates over the source counts, hence not included for the spectral analysis. Initially we have used an absorbed bremsstrahlung model, with a single Gaussian component for modelling the iron line complex, which could not be resolved by NuSTAR data. We obtain a good fit with $\chi^{2}/DOF=463(439)$. The column density of the absorber is $3.76_{-1.27}^{+1.32}\times10^{22}\;cm^{-2}$. This value is larger than the total galactic column density along the direction of the source,  $0.42\times10^{22}$cm$^{-2}$ \citep{HI4PI_2016}. This indicates a strong absorption specific to the source, is present in the spectra. The obtained continuum temperature is  $19.9_{-1.7}^{+1.9}$keV.  The Gaussian component shows a line energy of $6.59_{-0.05}^{+0.05}$ keV, with a $\sigma=247_{-59}^{+67}$eV, indicating presence of significant ionised Fe $K_{\alpha}$ lines.

Since the PSR is multitemperature in nature, the obtained bremsstrahlung temperature, indicates an average temperature of entire PSR. Therefore we used isobaric cooling flow model \texttt{mkcflow} \citep{Mushotzky_1988} to model the PSR \citep{Mukai_2003}. This model also incorporates the ionised line emissions including the ionised Fe K$_\alpha$ lines (He-like and H-like). We kept switch parameter set at 2 to determine the spectra based on \texttt{AtomDB 3.0.9} \footnote{\label{footsetting1} The same setting is used for subsequent spectral analysis, whenever required}. The redshift parameter is fixed at $1.35\times10^{-7}$ following the GAIA DR2 distance of $578\pm35$pc \citep{Gaia2}. We have also incorporated the absorption component \texttt{tbabs} . For the neutral iron K$\alpha$ line at 6.4 keV, we have added a Gaussian component with line central energy fixed at 6.4 keV and line width $\sigma$ fixed at 0 (as suggested by spectral analysis of XMM-Newton EPIC data, discussed in the next section). We obtained an acceptable fit with $\chi^{2}(DOF)=470(440)$. The overall absorber column density remains similar that of the bremsstrahlung fit within statistical uncertainty. The lower temperature reached the lower limit allowed by the model, so we fixed it at 0.0808keV. The upper temperature comes out to be $42.9_{-5.0}^{+5.5}$ keV. To check the presence of Compton reflection, we convolved \texttt{reflect} component  with the multitemperature cooling flow model, . It slightly improves the fit statistic to $\chi^{2}(DOF)=464(439)$, but the reflection amplitude, gives a high value $0.93_{-0.65}^{+0.87}$ with large uncertainty associated with it. The upper temperature is somewhat lowered in this case, standing at $27.9_{-5.1}^{+8.6}$keV. The value of the other parameters remain similar with no-reflection fit within statistical uncertainty. A strong and complex absorption can affect the spectra beyond 7 keV, thereby the continuum itself. Therefore, we can comment on reflection only after properly modelling the absorption in the source by including soft X-rays from XMM-Newton data.

\subsubsection{Phenomenological fit to XMM-Newton spectra}

\begin{figure}
	\includegraphics[width=\columnwidth, trim = {0.5cm 0.9cm 3.6cm 2.7cm}, clip]{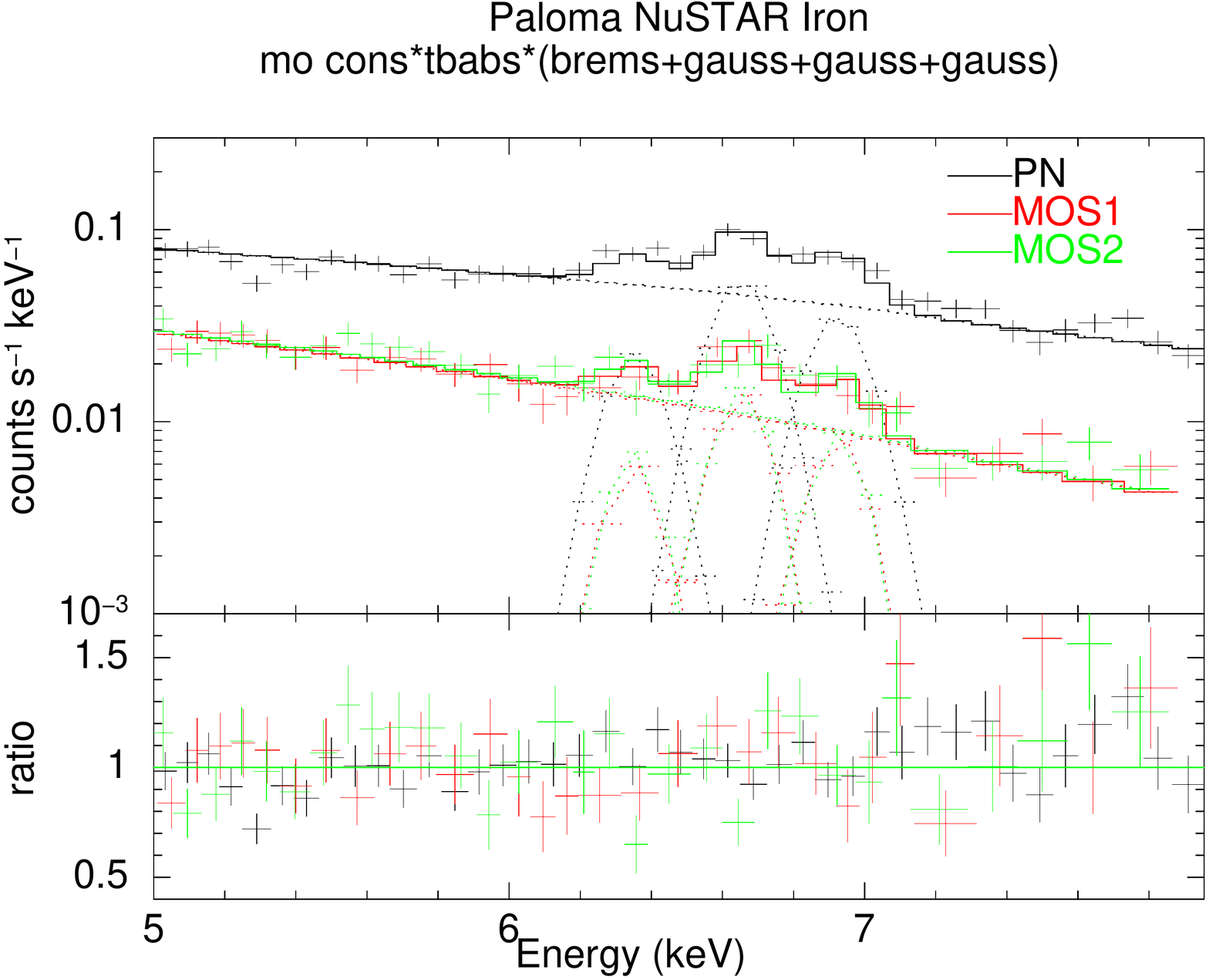}
    \caption{Fe K$_{\alpha}$ lines are fitted with three Gaussian components over an absorbed bremsstrahlung continuum using EPIC data in 5-8 keV range.  We show the spectra and data to model ratio in the \textcolor{blue}{top} and \textcolor{blue}{bottom} panel respectively}.
    \label{Fig:ironfig}
\end{figure}

First we have utilized the EPIC (PN, MOS1 and MOS2) spectra in 5-8 keV range to study the Fe K$_{\alpha}$ emission lines. We have used the absorbed bresmsstrahlung model to fit the underlying continuum. We have added three Gaussians to model the three iron K$_{\alpha}$ lines; neutral, He-like and H-like. The best-fit parameter values are reported in Table \ref{Tab:irontab} with the spectral fitting plotted in Fig. \ref{Fig:ironfig}. The line widths are kept fixed at 0 since all the Fe K$_\alpha$ lines are consistent with the instrumental resolution limit of EPIC ($\sim120$eV at 6.0keV). This shows that lines are narrow. We have fixed bremsstrahlung temperature and column density of absorber at $19.9$ keV and $3.76\times10^{22}\;cm^{-2}$, as obtained with NuSTAR only fit using bremsstrahlung model (see Sect. \ref{sec:nufit}), which otherwise couldn't be constrained. We notice the best-fit values of the line central energies are slightly redshifted, however, that is not statistically important as they agree with their expected theoretical values within error-bar.

\begin{table}

\renewcommand{\arraystretch}{1.2}
\caption{Probing Iron lines using EPIC data (5.0-8.0keV)}
\label{Tab:irontab}
\begin{tabular}{lcc}
\hline
Parameter                 & Unit            & Value                    \\ \hline
n$_{\text{H,tb}} $        & $10^{22} cm^{-2}$ & $3.76_f$                 \\
T$_{\text{br}}$           & keV             & $19.9_f$                 \\
N$_{\text{br}}$           & $\times10^{-3}$ & $ 1.82_{-0.05}^{+0.05} $ \\
E$_{\text{L, Neutal}}$   & keV             & $ 6.35_{-0.04}^{+0.05} $ \\
$\sigma_{\text{Neutal}}$  & eV              & $ 0_f $                  \\
eqw$_{\text{Neutal}}$     & eV              & $ 66_{-18}^{+30} $       \\
N$_{\text{L, Neutal}}$       & $\times10^{-5}$ & $ 0.7_{-0.2}^{+0.2} $    \\
E$_{\text{He-Like}}$  & keV             & $ 6.66_{-0.02}^{+0.02} $ \\
$\sigma_{\text{He-Like}}$ & eV              & $ 0_f $                  \\
eqw$_{\text{He-Like}}$    & eV              & $ 182_{-23}^{+28} $      \\
N$_{\text{He-Like}}$      & $\times10^{-5}$ & $ 1.9_{-0.2}^{+0.2} $    \\
E$_{\text{H-Like}}$   & keV             & $ 6.93_{-0.03}^{+0.02} $ \\
$\sigma_{\text{H-Like}}$  & eV              & $ 0_f $                  \\
eqw$_{\text{H-Like}}$     & eV              & $ 137_{-24}^{+24} $      \\
N$_{\text{H-Like}}$       & $\times10^{-5}$ & $  1.3_{-0.2}^{+0.2} $   \\ \hline
$\chi^{2}(DOF)$           &                 & $ 123(93) $              \\ \hline
\end{tabular}
\tablecomments{\# n$_{\text{H,tb}}$ represents column density of the absorber. T$_{\text{br}}$ and N$_{\text{br}}$ denote the temperature and normalisation of bremsstrahlung continuum. \\
$\star$ $E_L, \sigma, N_L$ and `eqw' respectively represent line central energy, line width, normalisation, and equivalent width of the corresponding Fe K$_{\alpha}$ line \\
$\ddagger$ $f$ denote the line parameter is fixed.}

\end{table}

\begin{figure}
	\includegraphics[width=\columnwidth, trim = {1.0cm 0.4cm 3.7cm 2.7cm}, clip]{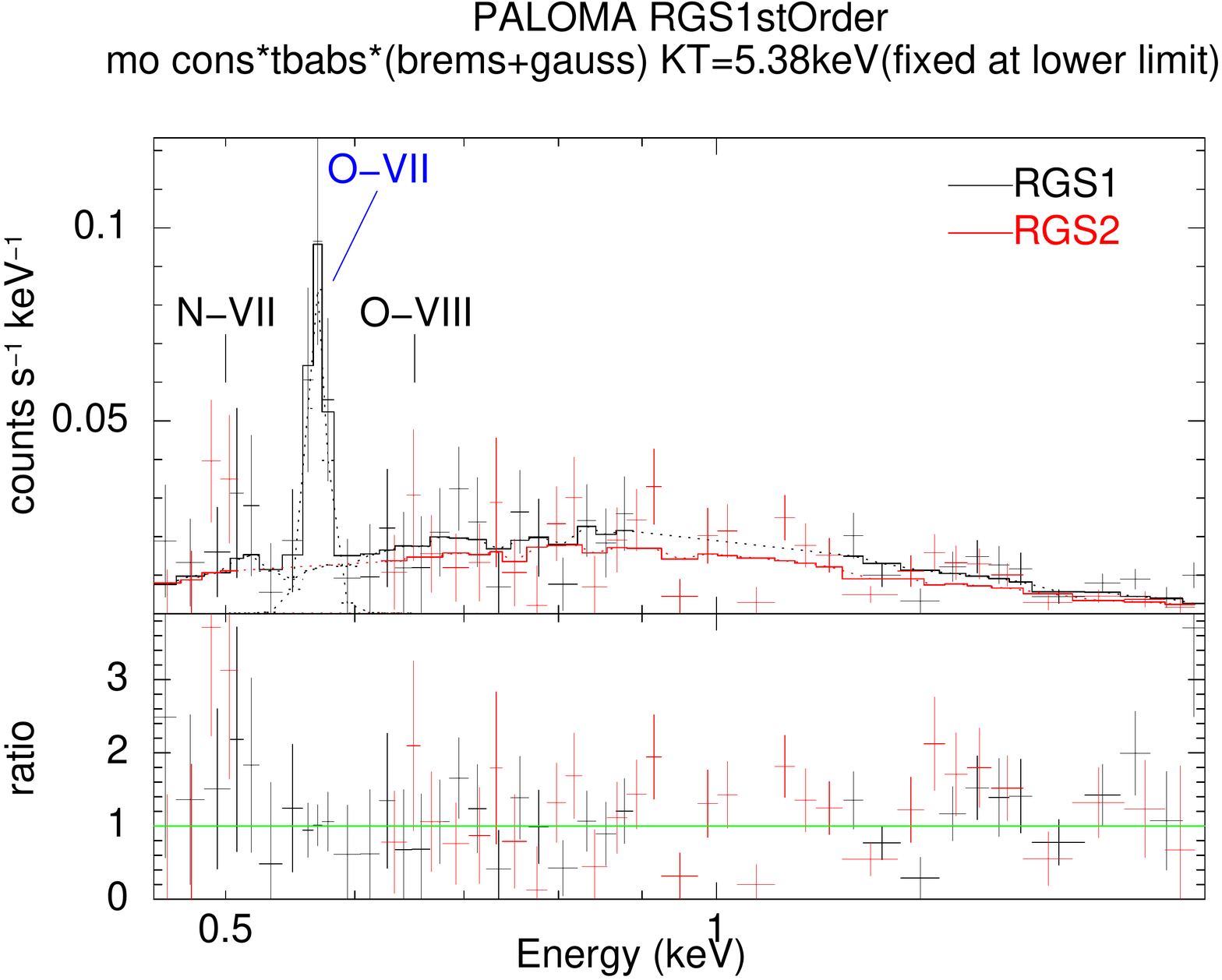}
    \caption{Fitting O-VII K$_{\alpha}$ line in RGS spectra. The spectral quality is poor, but O-VII line appears with significant strength and fitted with a Gaussian over an absorbed bremsstrahlung continuum. The spectra and data to model ratio are shown in \textcolor{blue}{top} and \textcolor{blue}{bottom} panel respectively. Expected location of O-VIII and N-VII K$_{\alpha}$ lines are marked.}
    \label{Fig:oxygenfig}
\end{figure}

We have next looked into the RGS spectra. The spectra from RGS1 and RGS2 are simultaneously fit in 0.45-2.0 keV. The signal to noise ratio of the spectral data is poor, with only 47\% of the total count is contributed from source. However, the ionised oxygen line (O-VII) in the spectra is clearly present. To get an overview of the ionised line, we fitted the line with a Gaussian component on top of an absorbed bremsstrahlung continuum. The spectra is shown in Fig. \ref{Fig:oxygenfig}. The line center appears at $0.568_{-0.004}^{+0.004}$keV with line width ($\sigma$) of $6_{-2}^{+5}$eV and an equivalent width of $117_{-77}^{+70}$eV. We note that, the O-VII line is composed of  fine atomic transition lines (intercombination, resonance and forbidden lines), which RGS can not resolve, thereby giving rise to a broad line width. There is visibly some excess residual around $0.5$ keV, expected from N-VII line. But poor signal to noise ratio for that line makes it statistically unimportant.

In order to constrain the absorption parameters, as well as the nature of the multitemperature PSR, the spectral data from all the three EPIC cameras were simultaneously fit in the 0.3-10.0 keV band. We have used ISM absorption model \texttt{tbabs} for modelling the overall absorption near the source as well as any intervening galactic absorption. The multitemperature plasma of the PSR has been modelled with the cooling flow model, implemented by \texttt{mkcflow}. The The neutral Fe K$_\alpha$ line is described by an narrow ($\sigma=0$) Gaussian component. The absorbed cooling flow model with a Gaussian component gives extremely poor fit statistic $\chi^{2}(DOF)=5447(424)$. Inclusion of an partial covering absorber dramatically improves the fit, with a $\chi^{2}(DOF)=604(421)$. The overall  column density appears to be $\sim6.3\times10^{20}\;cm^{-2}$ and partial covering absorber with column density $\sim6.08\times10^{22}\;cm^{-2}$ and covering fraction of $\sim65\%$. However, the upper temperature of the cooling flow model remains unconstrained (79 keV), the overall abundance is very large ($\sim 3.77$), and the fit statistic is still poor, and deserve further improvement. Inclusion of a \texttt{reflect} model substantially improves the fit to $\chi^{2}(DOF)=501(420)$, but it appear with an unphysical and large value of reflection amplitude $\sim8.24$. A poor fit with a single partial covering component suggests that absorption should require more elaborate modelling. Also, getting such a high value of reflection amplitude indirectly points that, absorption has an extended effect till the energy range beyond 7 keV which can also be mimicked by tail of reflection hump for an alleged strong Compton reflection. Therefore, instead of \texttt{reflect}, we introduced another partial covering absorber (\texttt{pcfabs$_1$*pcfabs$_2$}). Two partial covering absorbers mathematically describe abosorbers with three column densities and corresponding covering fraction (Eqn~\ref{Eqn:pcfabs})

\begin{equation}
\label{Eqn:pcfabs}
\small
\begin{split}
    [f_{pc1} e^{-n_{\text{H,pc1}}\sigma(E)}+(1-f_{pc1})]\times[f_{pc2} e^{-n_{\text{H,pc2}}\sigma(E)}+(1-f_{pc2})] \\
    = A + B e^{-n_{\text{H,pc1}}\sigma(E)} + C e^{-n_{\text{H,pc2}}\sigma(E)} + D e^{-(n_{\text{H,pc1}}+n_{\text{H,pc2}})\sigma(E)}
\end{split}
\end{equation}

Here $\sigma(E)$ stands for photoelectric cross section, $n_{\text{H,pc1}}$, $f_{pc1}$  and $n_{\text{H,pc2}}$, $f_{pc1}$ represent column densities and covering fraction contributed from each partial absorber model component. B, C, D are the functions of $f_{pc1}$ and $f_{pc2}$ and represent respective fraction of each of the resultant three column densities, and $A=1-(B+C+D)$.

The fit with two partial covering absorber improves the fit statistic significantly ($\chi^{2}(DOF)=478(419)$) in comparison to single partial partial absorber model. The column density of the overall absorber is $3.42_{-1.09}^{+1.13}\times 10^{20} \; cm^{-2}$. The column density and covering fraction of the two \texttt{pcfabs} components become $2.55_{-0.39}^{+0.46}\times10^{22}\;cm^{-2}$ with $65\pm3\%$ and $22.65_{-4.20}^{+5.40}\times10^{22}\;cm^{-2}$ with $50\pm5\%$. The upper temperature of the cooling flow model is constrained with a value of $43.2_{-12.2}^{+15.7}$keV. Though the error bars are large, it is expected due to limited coverage of hard X-rays by XMM-Newton data. The lower temperature of $0.86_{-0.24}^{+0.29}$ keV represents some mean temperature near the bottom region of the PSR. At this point we observe slight excess in the soft X-rays, possibly arising due to line emissions of low-Z and Fe L-shell emissions. The main excess is around $0.57$keV, representing the O-VII line, which is also observed from RGS spectra. To model that, we introduced an extra optically thin plasma emission component (\texttt{mekal}), appearing with a temperature $0.17_{-0.05}^{+0.04}$ keV, which decreased the fit statistic further to $\chi^{2}(DOF)=466(417)$. The corresponding F-statistic probability for inclusion of this extra component is $4.9\times10^{-3}$. 

It is evident that the source spectra involves a complicated absorption scenario, where it is paramount to model it appropriately. Motivated by the fact that single partial absorber is not enough to model the local intrinsic absorber, which indicated a distribution of column densities, we introduced \texttt{pwab} \citep{Done_1998} component to model the intrinsic absorber. This component assumes power law distribution of the covering fraction as a function of column density of neutral absorbers. The fit statistic is $\chi^{2}(DOF)=470(418)$, closely similar to the fit statistic obtained from previous two partial absorber fit. We kept the $n_{H,\text{min}}$ of the power law absorber fixed to $1\times10^{15}\;cm^{-2}$, which could not be constrained otherwise. The $n_{H,\text{max}}$ provides a value $32.09_{-6.78}^{+15.61}\times10^{22}\; cm^{-2}$ along with a power law index of $-0.60_{-0.01}^{+0.01}$. We could not constrain the column density of overall absorber either, which provided us with an upper limit of $<7.0\times10^{19}\;cm^{-2}$. However, The parameters from emission components remains similar with the fit of two partial covering absorber scenario within statistical uncertainty. Here we notice that, the F-test probability of introducing the extra optically thin component to model the emission features below 1 keV, is $1.5\times10^{-18}$, indicating much stronger necessity for this component than the two partial absorber case.

Given the complex intrinsic absorber scenario of Paloma, and emission features in soft X-rays, we have also performed an alternate modelling using \texttt{zxipab} \citep{Islam_2021}. This component can describe the absorption in the pre-shock flow as the power law distribution of warm photoionised absorber, hence providing an alternate description of the soft X-ray features. In this case, we do not introduce extra low temperature plasma emission component. We obtain a similarly good fit statistic like previous cases,  $\chi^{2}(DOF)=467(420)$. The overall absorber presents itself with a column density of $4.03_{-1.25}^{+1.42}\times 10^{20}\;cm^{-2}$, consistent with two partial absorber scenario. The warm absorber shows a column density value of $31.76_{-6.68}^{+10.25}\times10^{22}\;cm^{-2}$ for $n_{H,\text{max}}$, and we kept the $n_{H,\text{min}}$ frozen at the lowest value permissible for this parameter ($10^{15}\;cm^{-2}$), because of it becoming unconstrained.  The low temperature of cooling flow component now goes to the lower limit allowed by the model, so fixed at 0.0808 keV. The upper temperature ($34.5_{-7.3}^{+14.9}$keV) is consistent with the value obtained from previous two scenarios within statistical uncertainty.

\subsubsection{Simultaneous fit to broadband XMM-Newton and NuSTAR spectra}

Equipped with our understanding of the complex absorber and the overall continuum from the XMM-only and NuSTAR-only fit, we devised three model variants M$_1$, M$_2$, M$_3$ to model the broadband spectra in 0.3-40.0 keV. The description of the model variants are mentioned in Table \ref{Table:models}.

\begin{table*}
\centering
\renewcommand{\arraystretch}{1.0}
\caption{Model description in XSPEC notation }

\label{Table:models}
\begin{tabular}{@{}cc@{}}
\toprule
Abbreviation & Model                                    \\ \midrule
M1           & $\texttt{constant*tbabs*pcfabs*pcfabs*(mckflow+mekal+gauss)}$                    \\
M1-R         & $\texttt{constant*tbabs*pcfabs*pcfabs*(reflect*(mckflow+mekal)+gauss)}$ \\
M2           & $\texttt{constant*tbabs*pwab*(mckflow+mekal+gauss)}$                   \\
M2-R         & $\texttt{constant*tbabs*pwab*(reflect*(mckflow+mekal)+gauss)}$         \\
M3           & $\texttt{constant*tbabs*zxipab*(mckflow+gauss)}$                       \\
M3-R         & $\texttt{constant*tbabs*zxipab*(reflect*mckflow+gauss)}$               \\ \bottomrule
\end{tabular}
\end{table*}

\begin{figure}
	\includegraphics[width=1.\columnwidth, trim = {0.9cm 0.4cm 3.7cm 2.8cm}, clip]{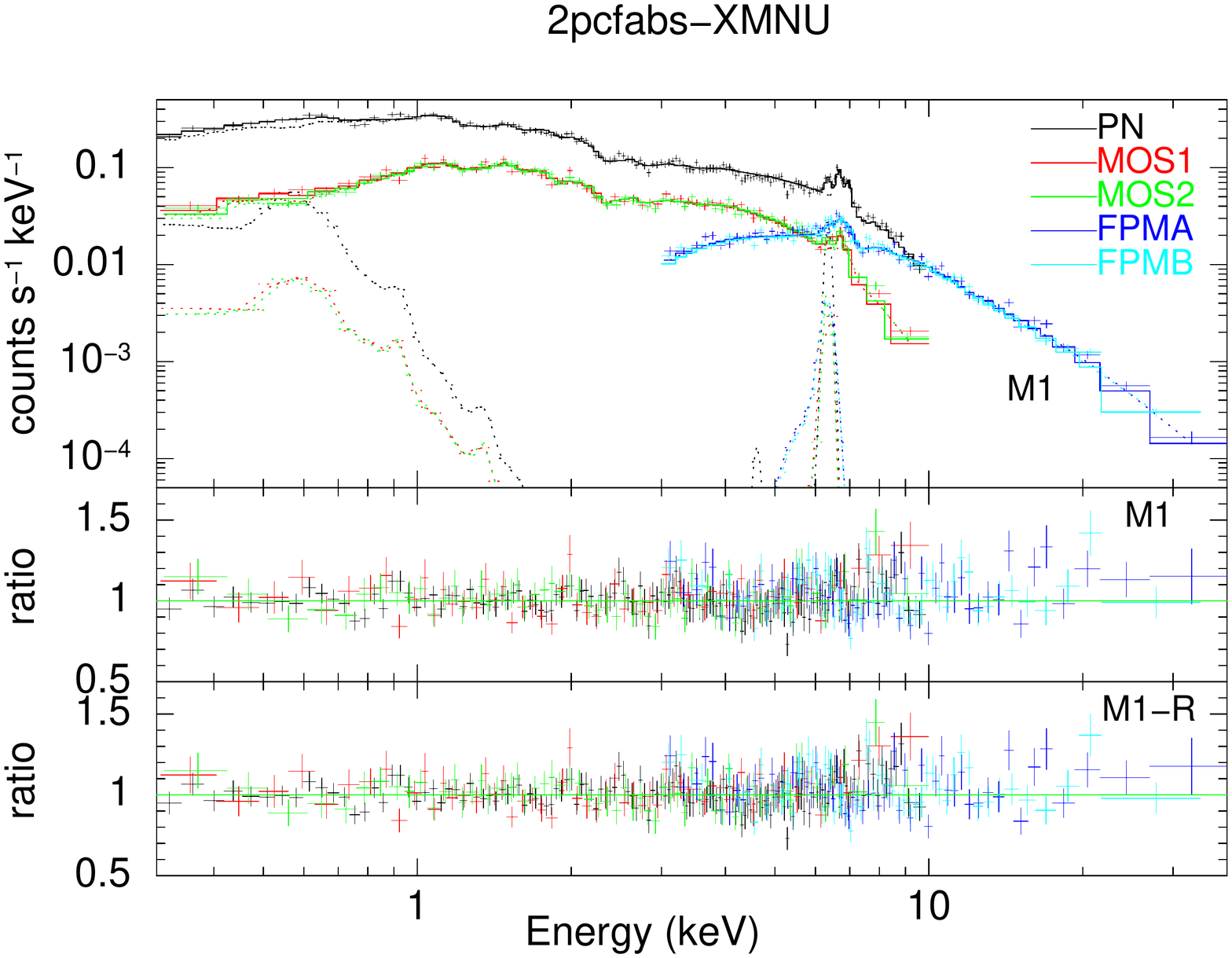}

    \caption{ The top and middle panel show the plots of the best-fit model and ratio (data/model) for the broadband spectra using model variant M1. We In the bottom panel we show the ratio plot using ``\texttt{reflect}" version of the model M1, i.e M1-R.}
    \label{Fig:avspectra}
\end{figure}

We have quoted the best fit values of the parameters in the Table \ref{Tab:avfit}. For representation, the spectral and ratio (data/model) plots using model M1 are shown in top and middle panels of Fig. \ref{Fig:avspectra}. The spectral parameters agree within the statistical uncertainty for all the three models. However, for model M2, we could not constrain the column density of overall absorber, and a upper limit ($<0.7\times10^{20}\; cm^{-2}$) is obtained. For a low value of overall absorber  which includes the galactic absorption, it maybe hard to constrain this parameter, given that we already have a complex intrinsic absorber component to describe the spectra. It is also to be noted that, the neutral power law distribution of absorber \texttt{pwab} and the ionised version \texttt{zxipab} have significant differences in predicting the model flux below $<0.8$ keV (Fig. 1 of \cite{Islam_2021}). This gives rise to differences in the obtained column densities of the overall absorber between model M2 and M3. All of the M1, M2, and M3 give us consistently statistically acceptable fit, with a $\chi^{2}(DOF)=952(859), 956(862), 952(863)$ respectively. Also, the emission parameters for all three model variants give statistically similar values.

\begin{table*}
\renewcommand{\arraystretch}{1.5}
\caption{Best-fit parameters using simultaneous broadband spectra in 0.3-40.0 keV }
\label{Tab:avfit}
\begin{tabular}{cccccccc}
\hline
\multicolumn{1}{c}{Parameter}         & Unit               & M1                       & M1-R                     & M2                        & M2-R                      & M3                       & M3-R                      \\ \hline \vspace*{-0.6cm} 
& & & & & & & \\ \hline
n$_{\text{H,tb}}$                      & 10$^{20}\;cm^{-2}$ & $ 4.5_{-1.2}^{+1.2} $    & $ 4.5_{-1.2}^{+1.2} $    & $ <0.7 $                  & $ <0.7 $                  & $ 4.7_{-1.3}^{+1.3} $    & $ 4.5_{-1.3}^{+1.4} $     \\
n$_{\text{H,pc1}}$                     & 10$^{22}\;cm^{-2}$ & $ 2.2_{-0.4}^{+0.4} $    & $ 2.1_{-0.2}^{+0.4} $    & $ ... $                   & $ ... $                   & $ ... $                  & $ ... $                   \\
pcf1/ log(xi)$^{\dagger\;a}$                          &                    & $ 0.66_{-0.03}^{+0.02} $ & $ 0.66_{-0.02}^{-0.02} $ & $ ... $                   & $ ... $                   & $ 0.6_{-0.1}^{+0.1} $     & $ 0.6_{-0.1}^{+0.1} $      \\
n$_{\text{H,pc2}}$/ n$_{\text{H,max}}$ $^{\dagger\;b}$ & 10$^{22}\;cm^{-2}$ & $ 20.0_{-3.1}^{+3.1} $   & $ 19.0_{-2.8}^{+3.2} $   & $27.6 _{-4.1}^{+7.3} $    & $ 25.2_{-4.1}^{+6.8} $    & $ 30.5_{-5.4}^{+7.3} $   & $ 28.8_{-4.6}^{+7.9} $    \\
pcf2/ $\beta$ $^{\dagger\;c}$                          &                    & $ 0.54_{-0.04}^{+0.03} $ & $ 0.54_{-0.03}^{+0.03} $ & $ -0.59_{-0.01}^{+0.01} $ & $ -0.58_{-0.02}^{+0.01} $ & $ -0.55_{-0.01}^{+0.01} $ & $ -0.58_{-0.02}^{+0.02} $ \\
T$_{me}$                               & keV                & $ 0.18_{-0.05}^{+0.03} $ & $ 0.18_{-0.05}^{+0.03} $ & $ 0.19_{-0.02}^{+0.02} $  & $ 0.19_{-0.02}^{+0.02} $  & $ ... $                  & $ ... $                   \\
N$_{me}$                               & $ 10^{-4} $        & $ 1.7_{-0.1}^{+1.1} $    & $ 2.0_{-0.9}^{+1.2} $    & $ 4.5_{-1.0}^{+1.7} $     & $ 5.5_{-1.3}^{+1.5} $     & $ ... $                  & $ ... $                   \\
T$_{1,mkc}$                            & keV                & $ 0.67_{-0.17}^{+0.18} $ & $ 0.67_{-0.16}^{+0.19} $ & $0.68 _{-0.19}^{+0.17} $  & $0.69 _{-0.17}^{+0.18} $  & $ 0.0808_f $             & $ 0.0808_f $              \\
T$_{2, mkc}$                           & keV                & $ 34.2_{-2.5}^{+4.2} $   & $ 29.6_{-4.9}^{+4.9} $   & $ 33.6_{-4.2}^{+1.9} $    & $ 27.9_{-3.9}^{+5.8} $    & $  31.7_{-3.5}^{+3.3} $  & $ 28.5_{-3.6}^{+5.4} $    \\
N$_{mkc}$                              & $ 10^{-11} $       & $ 11.1_{-1.3}^{+1.0} $   & $ 12.0_{-1.4}^{+1.5} $   & $ 10.6_{-1.1}^{+1.9} $    & $ 11.8_{-1.5}^{+1.6} $    & $ 11.8_{-1.4}^{+1.9} $   & $ 12.3_{-1.8}^{+1.7} $    \\
Z                                      & $ Z_{\odot} $      & $ 1.4_{-0.2}^{+0.2} $    & $ 1.2_{-0.2}^{+0.2} $    & $ 1.2_{-0.2}^{+0.2} $     & $ 1.1_{-0.2}^{+0.2} $     & $ 1.2_{-0.2}^{+0.2} $    & $ 1.1_{-0.2}^{+0.2} $     \\
$\Omega/2\pi$                          &                    & $ ... $                  & $ 0.39_{-0.25}^{+0.48} $ & $ ... $                   & $ 0.47_{-0.35}^{+0.43} $   & $ ... $                  & $ 0.26_{-0.26}^{+0.45} $  \\
E$_{\text{L}}$                                  & keV                & $ 6.36_{-0.02}^{+0.04} $ & $ 6.36_{-0.04}^{+0.03} $ & $ 6.36_{-0.04}^{+0.04} $  & $ 6.37_{-0.05}^{+0.04} $  & $ 6.34_{-0.01}^{+0.06} $ & $ 6.36_{-0.04}^{+0.04} $  \\
$\sigma$                               & eV                 & $ 0_{f} $                & $ 0_{f} $                & $ 0_{f} $                 & $ 0_{f} $                 & $ 0_{f} $                & $ 0_{f} $                 \\
N$_{\text{L}}$                            & $ 10^{-5} $        & $ 1.0_{-0.2}^{+0.2} $    & $ 0.9_{-0.2}^{+0.2} $    & $ 0.9_{-0.2}^{+0.2} $     & $ 0.8_{-0.2}^{+0.2} $     & $ 0.9_{-0.2}^{+0.2} $    & $ 0.8_{-0.2}^{+0.2} $     \\ \hline
$\chi^{2}(DOF)$                        &                    & $ 952(859) $             & $ 948(858) $             & $ 956(862) $              & $ 952(861) $              & $ 952(863) $             & $ 951(862)$               \\ \hline
\end{tabular}
\tablecomments{$\dagger a$: Read pcf1 (covering fraction for first partial absorber) for model M1, and log(xi) (ionisation parameter of warm power law absorber) for model M3. \\
$\dagger b$ Read n$_{\text{H,pc2}}$ (column density of second partial absorber) for model M1, and n$_{\text{H,max}}$ (maximum column density of the neutral (and warm) power law absorber) for model M2 (and  M3). \\
$\dagger c$ Read pcf2 (covering fraction for second partial absorber) for model M1, and $\beta$ (power law index of neutral (and warm) absorber)  for model M2 (and M3).\\
\# $n_{H,\text{tb}}$ and $n_{H,\text{pc1}}$ respectively denote column density of overall and first partial absorber. T$_{me}$, N$_{me}$ are temperature and normalisation of optically thin cold plasma emission component. T$_{1, mkc}$, T$_{2, mkc}$ and N$_{mkc}$ denote upper \& lower temperature  and normalisation of isobaric cooling flow component. Z stands for overall abundance. E$_{\text{L}}$, $\sigma$, N$_{\text{L}}$ represent line energy, line width and normalization of the Gaussian component for the neutral Fe K$_{\alpha}$ line respectively. \\
$\star$ $f$ denotes the parameter is fixed.}

\end{table*}

Once our spectral model has satisfactorily constrained the absorption parameters and the continuum, we attempted to find the contribution of Compton reflection. We adopted model variants M1-R, M2-R and M3-R (for model description, see Table \ref{Table:models}), with the abundance parameters of \texttt{reflect} component linked to that of emission components. We used default value of the inclination angle ($i$), set at $\mu=\cos{i}=0.45$. The three models produced a reflection amplitude of $0.39_{-0.25}^{+0.48} $, $ 0.47_{-0.35}^{+0.43} $ and $ 0.26_{-0.26}^{+0.45} $ along with an improvement in fit-statistic, however not large, of $\Delta\chi^{2}(\Delta DOF)=4(1), 4(1)$ and $1(1)$ respectively (Table \ref{Tab:avfit}). The ratio (data/model) plot using model M1-R  is shown in bottom panel of Fig. \ref{Fig:avspectra}. Though mutually agreeing values of reflection amplitudes are obtained in all three models, the large statistical uncertainty and small improvement in fit-stat puts us in a delicate situation to comment about its ``robust" presence in our spectra.

\subsubsection{Phase resolved spectroscopy}

Our orbit folded lightcurve (Fig. \ref{Fig:foldedlightcurve}) showed prominent variation in the HR-1, indicating the presence of the change in the complex intrinsic absorber.  We extracted phase-resolved spectra in the four phase intervals as defined in section \ref{Sect:Folded Lightcurves}. We have quoted the best-fit parameters for the phase-resolved spectra in Table \ref{Tab:phaseresolve}. We kept the column density of the overall absorber fixed at the best fit value from the phase averaged spectral fitting. The temperatures of the plasma emission components remain same within statistical uncertainty for all the phases as well as with the phase-averaged value, for any particular model variant. Though the column density of the complex absorbers also remain same within error-bar, our spectral fitting clearly reveals the change in the covering fraction parameters. For phase (a) the model M1 show highest value of the covering fraction (pcf1 and pcf2) for the intrinsic absorbers  ($0.79_{-0.04}^{+0.04}$ and $0.73_{-0.06}^{+0.05}$), whereas for the phase (c), the lowest values of those parameters are obtained ($ 0.49_{+0.06}^{-0.11}$ and $0.31_{-0.16}^{+0.12}$). The parameter values (pcf1 and pcf2) for other two phases - phase (b) ($0.67_{-0.04}^{+0.05}$ and $0.50_{-0.06}^{+0.05}$) and phase (d) ($0.70_{-0.04}^{+0.03}$ and $0.54_{-0.06}^{+0.04}$) are consistent with each other. The model M2 and M3 follow same trend with the power law index for the complex absorber. We obtain highest value of the power law index($\sim-0.3$) for phase (a) and lowest value ($\sim-0.7$) for phase (c), whereas the phase (b) and (d) gives intermediate values ($\sim -0.5$), similar to each other. 
For model $M_3$, we also obtain a low value of ionisation parameter ($\sim -1.3$) during phase (c), compared to other three phases ($\sim 0.7, 0.7, 0.5$ for (a), (b), (d) respectively).

\begin{table*}
\movetabledown=4.2in
\begin{rotatetable*}
\small
\renewcommand{\arraystretch}{1.5}
\caption{Best fit parameters for phase Resolved spectra in 0.3-40.0 keV }
\label{Tab:phaseresolve}

\begin{tabular}{@{\extracolsep{0.1\tabcolsep}}c@{\hspace{0.1\tabcolsep}}cc@{\hspace{-4.0\tabcolsep}}c@{\hspace{-4.0\tabcolsep}}c@{\hspace{-4.0\tabcolsep}}cc@{\hspace{-4.0\tabcolsep}}c@{\hspace{-4.0\tabcolsep}}c@{\hspace{-4.0\tabcolsep}}cc@{\hspace{-4.0\tabcolsep}}c@{\hspace{-4.0\tabcolsep}}c@{\hspace{-4.0\tabcolsep}}c@{}}
\toprule
Parameter                              & Unit               & \multicolumn{4}{c}{M1}                                                                                    & \multicolumn{4}{c}{M2}                                                                                        & \multicolumn{4}{c}{M3}                                                                                        \\ \cmidrule(l){3-6} \cmidrule(l){7-10} \cmidrule(l){11-14}
&                    & \begin{tabular}[c]{@{}c@{}}(a)\\ 0.10-0.22\end{tabular} & \begin{tabular}[c]{@{}c@{}}(b)\\ 0.22-0.50\end{tabular} & \begin{tabular}[c]{@{}c@{}}(c)\\ 0.50-0.80\end{tabular} & \begin{tabular}[c]{@{}c@{}}(d)\\ 0.80-1.10\end{tabular} & \begin{tabular}[c]{@{}c@{}}(a)\\ 0.10-0.22\end{tabular} & \begin{tabular}[c]{@{}c@{}}(b)\\ 0.22-0.50\end{tabular} & \begin{tabular}[c]{@{}c@{}}(c)\\ 0.50-0.80\end{tabular} & \begin{tabular}[c]{@{}c@{}}(d)\\ 0.80-1.10\end{tabular} & \begin{tabular}[c]{@{}c@{}}(a)\\ 0.10-0.22\end{tabular} & \begin{tabular}[c]{@{}c@{}}(b)\\ 0.22-0.50\end{tabular} & \begin{tabular}[c]{@{}c@{}}(c)\\ 0.50-0.80\end{tabular} & \begin{tabular}[c]{@{}c@{}}(d)\\ 0.80-1.10\end{tabular}                 \\ \cmidrule(l){1-2} \cmidrule(l){3-6} \cmidrule(l){7-10} \cmidrule(l){11-14}
n$_{\text{H,tb}}$                      & 10$^{20}\;cm^{-2}$ & $ 4.5_{f} $              & $ 4.5_{f} $              & $ 4.5_{f} $              & $ 4.5_{f} $              & $ 0.7_f $                 & $ 0.7_f $                 & $ 0.7_f $                 & $ 0.7_f $                 & $ 4.7_{f} $               & $ 4.7_{f} $               & $ 4.7_{f} $               & $ 4.7_{f} $               \\
n$_{\text{H,pc1}}$                     & 10$^{22}\;cm^{-2}$ & $ 2.3_{-0.8}^{+0.7} $    & $ 1.8_{-0.3}^{+0.5} $    & $ 2.9_{-1.5}^{+1.4} $    & $ 1.9_{-0.5}^{+0.4} $    & $ ... $                   & $ ... $                   & $ ... $                   & $ ... $                   & $ ... $                   & $ ... $                   & $ ... $                   & $ ... $                   \\
pcf1/ log(xi)                          &                    & $ 0.79_{-0.04}^{+0.04} $ & $ 0.67_{-0.04}^{+0.05} $ & $ 0.49_{-0.11}^{+0.06} $ & $ 0.70_{-0.04}^{+0.03} $ & $ ... $                   & $ ... $                   & $ ... $                   & $ ... $                   & $ 0.66_{-0.16}^{+0.16} $  & $ 0.68_{-0.12}^{+0.12} $  & $ -1.30_{-0.19}^{+0.21} $ & $ 0.46_{-0.09}^{+0.10} $  \\
n$_{\text{H,pc2}}$/ n$_{\text{H,max}}$ & 10$^{22}\;cm^{-2}$ & $ 20.9_{-4.3}^{+5.6} $   & $ 16.3_{-4.3}^{5.7} $    & $ 21.2_{-20.2}^{+21.5} $ & $ 18.3_{-4.1}^{+5.4} $   & $ 34.6_{-10.0}^{+17.7} $  & $ 18.1_{-4.7}^{+9.3} $    & $ 20.1_{-8.8}^{+15.3} $   & $ 21.7_{-5.7}^{+10.7} $   & $ 38.9_{-10.2}^{+17.0} $  & $ 20.2_{-4.3}^{+7.0} $    & $ 18.6_{-10.2}^{+39.1} $  & $ 22.8_{-4.8}^{+9.2} $    \\
pcf2/ $\beta$                          &                    & $ 0.73_{-0.06}^{+0.05} $ & $ 0.50_{-0.06}^{+0.05} $ & $ 0.31_{-0.16}^{+0.12} $ & $ 0.54_{-0.06}^{+0.04} $ & $ -0.36_{-0.02}^{+0.04} $ & $ -0.59_{-0.03}^{+0.03} $ & $ -0.74_{-0.03}^{+0.03} $ & $ -0.57_{-0.02}^{+0.02} $ & $ -0.32_{-0.04}^{+0.04} $ & $ -0.55_{-0.02}^{+0.02} $ & $ -0.77_{-0.02}^{+0.03} $ & $ -0.52_{-0.02}^{+0.02} $ \\
T$_{me}$                               & keV                & $ <0.27 $                & $ <0.19 $                & $ 0.18_f $               & $ 0.12_{-0.04}^{+0.05} $ & $ 0.18_{-0.08}^{+0.04} $  & $ 0.17_{-0.03}^{+0.03} $  & $ <0.22_{-0.06}^{+0.05} $ & $ 0.12_{-0.02}^{+0.04} $  & $ ... $                   & $ ... $                   & $ ... $                   & $ ... $                   \\
N$_{me}$                               & $ 10^{-4} $        & $ 4.4_{-3.7}^{+4.5} $    & $ 3.4_{-1.6}^{+2.1} $    & $ 0.4_{-0.4}^{+0.7} $    & $ 3.7_{-2.6}^{+5.8} $    & $ 16.8_{-8.6}^{+23.2} $   & $ 5.2_{-1.7}^{+3.2} $     & $ 1.3_{-0.6}^{+0.9} $     & $ 8.0_{-3.4}^{+7.8} $     & $ ... $                   & $ ... $                   & $ ... $                   & $ ... $                   \\
T$_{1,mkc}$                            & keV                & $ <1.28 $                & $ <0.98 $                & $ <0.90 $                & $ <0.86 $                & $ <1.41 $                 & $ 0.72_{-0.35}^{+0.53} $  & $ <0.87 $                 & $ 0.58_{-0.24}^{+0.30} $  & $0.0808_f $               & $0.0808_f $               & $0.0808_f $               & $0.0808_f $               \\
T$_{2, mkc}$                           & keV                & $ 41.2_{-10.4}^{+17.5} $ & $ 35.4_{-6.2}^{+5.7} $   & $ 38.2_{-9.3}^{+13.9} $  & $ 43.6_{-7.4}^{+11.2} $  & $ 35.7_{-10.8}^{+17.6} $  & $ 36.8_{-7.7}^{+9.4} $    & $ 38.0_{-10.0}^{+13.0} $  & $ 44.2_{-8.6}^{+11.1} $   & $ 33.3_{-10.2}^{+17.0} $  & $ 33.7_{-6.1}^{+9.6} $    & $ 28.3_{-16.0}^{+11.6}$    & $ 40.3_{-8.3}^{+9.0} $    \\
N$_{mkc}$                              & $ 10^{-11} $       & $ 12.1_{-3.7}^{+4.2} $   & $ 12.0_{-2.3}^{+2.8} $   & $ 4.6_{-1.2}^{+1.9} $    & $ 10.9_{-2.0}^{+2.4} $   & $ 13.1_{-4.5}^{+8.6} $    & $ 10.6_{-2.1}^{+3.6} $    & $ 4.4_{-0.1}^{+0.2} $     & $ 9.9_{-2.0}^{+3.4} $     & $ 14.9_{-4.8}^{+9.2} $    & $ 11.8_{-2.6}^{+3.2} $    & $ 6.5_{-1.7}^{+2.8}$   & $ 11.1_{-2.1}^{+3.6} $    \\
Z                                      & $ Z_{\odot} $      & $ 1.6_{-0.6}^{+0.6} $    & $ 1.8_{-0.4}^{+0.5} $    & $1.3_{-0.4}^{+0.6} $     & $1.6_{-0.4}^{+0.7}  $    & $ 1.2_{-0.6}^{+0.8} $     & $ 1.7_{-0.4}^{+0.6} $     & $ 1.3_{-0.4}^{+0.6} $     & $ 1.5_{-0.4}^{+0.5} $     & $ 1.2_{-0.5}^{+0.6} $     & $ 1.6_{-0.4}^{+0.5} $     & $ 0.8_{-0.3}^{+0.3} $     & $ 1.3_{-0.4}^{+0.4} $     \\
LineE                                  & keV                & $ 6.48_{-0.12}^{+0.08} $ & $ 6.32_{-0.03}^{+0.03} $ & $ 6.44_{-0.10}^{+0.07} $ & $ 6.38_{-0.05}^{+0.05} $ & $ 6.48_{-0.08}^{+0.09} $  & $ 6.32_{-0.03}^{+0.03} $  & $ 6.43_{-0.09}^{+0.07} $  & $ 6.38_{-0.05}^{+0.06} $  & $ 6.48_{-0.07}^{+0.09} $  & $ 6.32_{-0.03}^{+0.04} $  & $ 6.45_{-0.05}^{+0.10} $  & $ 6.38_{-0.05}^{+0.05} $  \\
$\sigma$                               & eV                 & $ 0_f $                  & $ 0_f $                  & $ 0_f $                  & $ 0_f $                  & $ 0_f $                   & $ 0_f $                   & $ 0_f $                   & $ 0_f $                   & $ 0_f $                   & $ 0_f $                   & $ 0_f $                   & $ 0_f $                   \\
N$_{line}$                             & $ 10^{-5} $        & $  1.5_{-0.7}^{+0.8} $   & $ 1.6_{-0.4}^{+0.4} $    & $ 0.5_{-0.2}^{+0.3} $    & $ 1.1_{-0.4}^{+0.4} $    & $ 1.4_{-0.7}^{+0.8} $     & $ 1.4_{-0.4}^{+0.4} $     & $ 0.5_{-0.2}^{+0.2} $     & $ 1.0_{-0.3}^{+0.4} $     & $1.3_{-0.7}^{+0.8}$       & $ 1.4_{-0.4}^{+0.4} $      & $ 0.5_{-0.3}^{+0.3} $     & $ 10.1_{-3.5}^{+3.4} $     \\ \cmidrule(l){1-2} \cmidrule(l){3-6} \cmidrule(l){7-10} \cmidrule(l){11-14}
$\chi^{2}(DOF)$                        &                    & $ 200(203) $             & $ 393(427) $             & $ 347(304) $             & $ 503(463) $             & $ 203(205) $              & $ 399(429) $              & $ 345(305) $              & $ 506(465) $              & 210(207)                  & $ 405(431) $              & $ 355(307) $              & $ 502(467) $              \\ \bottomrule
\end{tabular}
\tablecomments{\hspace{3cm} Symbols carry same meaning as that of Table \ref{Tab:avfit}}
\end{rotatetable*}
\end{table*}

\section{Discussion}
\label{Sec:discussion}

Paloma, one of a unique sources among IPs, has been observed simultaneously in the broadband X-rays (0.3-40 keV) for the first time. We have studied the X-rays properties of the source and in the following section we discuss our results.

\subsection{Modulations in the lightcurve}
\label{Sec:discussion:modulation}
 The single hump like structure in the orbit folded lightcurve covering a full cycle denotes we are seeing emissions from at least one pole at any point of time during the orbital motion of the WD. The energy dependent dips arise due to  presence of complex inhomogeneous absorber. During dip, around the phase (a) ($\phi=0.10-0.22$), we believe the column density of the absorber is the maximum along the line of sight, causing the soft X-rays below 3 keV to undergo maximum absorption. This is what causes the peak in the HR1 (Fig \ref{Fig:foldedlightcurve} )at same phase. This absorption feature also appears in the 3-10 keV band, indicating that the column density of the absorber is very high during this phase. However, the amplitude of the dip decreases with increasing energy. The complex intrinsic absorber also describes another extra dip at $\phi \sim 0.5$ in the soft X-rays below 3 keV, but does not impact the lightcurve above 3 keV.
 
The periodic absorption features present at $\phi \sim$ 0.16 and 0.5 in the orbital cycle for the soft to medium X-rays (0.3-10.0 keV) of the PN lightcurve can be contributed from absorbing material produced by stream-disc or stream-magnetosphere interactions \citep{Norton_1996, Parker_2005}. However, due to close proximity of the orbital and spin periods, and short duration of the PN lightcurve (covering few spin and orbital cycles), the periodic absorption features from accretion curtain or accretion stream arising in the spin cycle can also contribute to the features in orbit folded PN lightcurve.

The significant power at orbital frequency $\Omega$ is representative of strong periodic variation at this period. The minima at $\phi\sim0.7$ during phase (c) is present in all the energy bands in the orbit folded lightcurves, obtained from both the telescopes. Presence of dense blob of material fixed in orbital frame along the line of sight, capable of obstructing hard X-rays can be a possible explanation for this feature during phase (c). We also note relatively weaker peak at $\Omega$ in the soft 0.3-3.0 keV band w.r.t to the medium 3.0-10.0 keV band. This can happen because of the features coming from the absorbers fixed in spin phase affect the soft X-rays (0.3-3.0 keV) in PN lightcurve more dominantly, and gets superposed with the modulation features arising in the orbital cycle, thereby reducing the power at $\Omega$ in the soft energy band.

The earlier photometric observations revealed variation of spin modulations over a complete beat cycle. \citet{Littlefield_2022} found that spin profile changes from single peak to double peak structure and they proposed pole switching or grazing eclipse. \citet{Schwarz_2007} showed the change in modulations of the double hump structure (Their case B). They argued for pole shifting scenario. These kind of variations in spin modulations over a beat cycle weakens the power at spin frequency in the NuSTAR power spectra (both 3-10 and 10-40 keV energy band), derived from more than a beat-period long lightcurve. In pole switching case the power at the frequency peak can be shifted from $\omega$ to $\Omega$ and $2\omega-\Omega$ approximately equally \citep{Littlefield_2019}. But in the NuSTAR power spectra we notice the $2\omega-\Omega$ peak is even weaker than $\omega$. Therefore, for a pole switching scenario, if some power is transferred from $\omega$ to $\Omega$ in the NuSTAR power spectra, then that power shouldn't be of substantial amount. It indicates that the strong orbital frequency peak is a distinguishing feature for Paloma, irrespective of weaker spin frequency peak. However, this is not commonplace in IPs, for e.g in a systematic study of several IPs by \citet{Parker_2005}, orbital peaks appear to be weaker than or comparable with spin peaks.

\subsection{Nature of the intrinsic absorber and the variation with orbital phase}

The inherently complex nature of the intrinsic absorber is supported from spectral analysis. The simplistic scenario, implemented using two multiplicative partial covering model components as described by model M1, mathematically represents intrinsic inhomogeneous absorbers with at least three distinct column densities with their respective fractional contribution. From the fit to the phase averaged spectra, we calculate three such column densities as $2.2\times10^{22} cm^{-2}$, $20.0\times10^{22} cm^{-2}$ and $22.2\times10^{22} cm^{-2}$ appearing with fractional contribution of $0.30, 0.18$, and $0.35$ (following Eqn. \ref{Eqn:pcfabs}). 
We can physically explain this as the X-ray emission encounters the intrinsic absorbers which are mentioned in earlier susbection (Sect. \ref{Sec:discussion:modulation}). These absorbers are inhomogenous and do not have a unique column density rather a distribution of column densities.  This scenario is supported by model M2 and M3, where we assume power law distribution of the column densities of the intrinsic absorber. The maximum values of the column density are, $\sim28\times10^{22}\;cm^{-2}$, consistent with model M1.

During orbital phase (a), the high value of the covering fractions for model M1, or high values of the power law indices ($\beta$) of the complex absorber for model M2, M3 agree with the sharp dip that we see in the folded lightcurve. This sharp excess in absorption is contributed when a large part of emission encounters the absorber. During the other two phases (b) and (d), we obtain similar value of covering fraction (model M1) or the $\beta$ of the complex absorber (model M2 and M3). This indicates the situation while lesser and lesser amount of emission pass through the absorber, as the system rotates and during phase (c), the least value of the covering fractions (model M1) or $\beta$ (model M2 and M3) are obtained.

\subsection{Iron K$_\alpha$ lines and Compton reflection}

The XMM-Newton EPIC spectra in Fig \ref{Fig:ironfig} presents three Fe K$_{\alpha}$ line complex, where the He-like Fe K$_\alpha$ line appear with maximum strength (equivalent width $182_{-23}^{+28}$eV), and the neutral line is the weakest one (equivalent width $66_{-18}^{+30}$eV). All the three lines are narrow, with line width being consistent with instrument resolution.

The contribution of the neutral Fe K$_\alpha$ line comes from both the intrinsic absorber, as well as from the Compton reflection of the X-rays from the WD surface \citep{Ishida_1999}. A strong and complex absorber, with a weak neutral Fe K$_\alpha$ line suggests that the absorber is likely to be the main contributor for the strength of this line for Paloma, however, there is a possibility that some amount is contributed via Compton reflection.

In the broadband fitting, The "\texttt{reflect}" versions of the models (i.e M1-R, M2-R, M3-R) display small improvement in fit-statistic. This suggests that reflection is weak in the source.

\begin{figure}
	\includegraphics[width=\columnwidth, trim = {0.9cm 0.4cm 3.7cm 2.5cm}, clip]{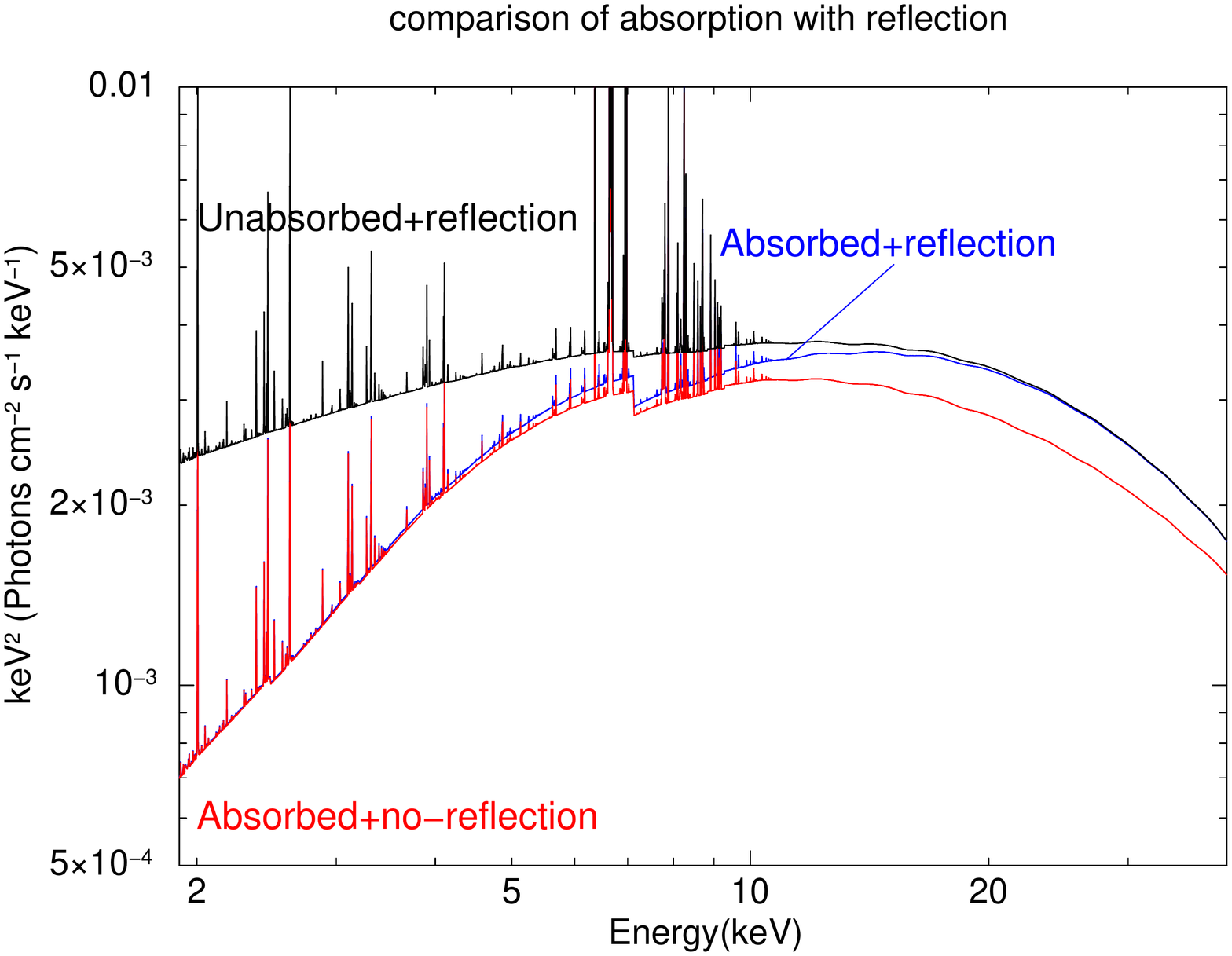}
    \caption{Effect of complex absorber and Compton reflection in the spectral modelling. Spectral parameters from model M1-R have been used. The blue curve represents the model spectrum including both reflection and absorption. For the black curve, the absorption is set to zero, but reflection is present. The red curve denotes the spectra when absorption is present but reflection is not. }
    \label{Fig:absrefl}
\end{figure}

In order to check the degree of degeneracy between absorption and reflection, we plotted (Fig. \ref{Fig:absrefl}) the model spectra (using best-fit values of parameters of M1-R fit, obtained in Table \ref{Tab:avfit}) when both of the absorption and reflection are present and when either one of them is absent. This plot shows that the strong intrinsic absorber in Paloma have effect well around 15 keV (where black and blue curves merge) whereas, the lower end of the reflection hump tail is comfortably extended till 7 keV (where red and blue curves merge).  It indicates the degeneracy of the reflection and absorption in this energy range, which can be eliminated by securely detecting the reflection hump beyond 15 keV. The model spectra show a reflection hump feature in 10-30 keV which we could not robustly detect it in the observed spectra as the improvement in fit-statistics is minimal. To be noted, \cite{Mukai_2015} have detected strong reflections in IPs. So, reflection in IPs is an important physical process and can be studied in detail provided the spectral quality is such that the detection is robust.

The reflection amplitude ($\Omega/2\pi$) physically denotes the shock height (h) over the surface the WD. Theoretically, reflection amplitude of 1 denotes the shock is happening just over the WD surface whereas a reflection amplitude 0 denotes a tall shock \citep{Mukai_2017}.  If we consider that there is certain contribution of Compton reflection in Paloma, though we agree that the statistical improvement to fit is small, we obtain physically acceptable values of the reflection amplitudes $\sim0.39$, $0.47$ and $0.26$ from models M1-R, M2-R and M3-R respectively. These values are much less than the value calculated by \cite{Joshi_2016} ($\sim 2.4$) using XMM-Newton only data. However, remembering the large statistical uncertainty associated with $\Omega/2\pi$ for any of our model variants, we refrain from deducing quantitative information, like value of the shock-height, based on that.

\subsection{Case of Collisionally excited plasma and photoionised absorber}

The RGS spectra shows signature of a strong O-VII line. Due to the poor quality of the data, we could not detect any other lines with statistical significance. Agreeing with RGS spectra, the ionised oxygen line also appears as a small hump around 0.57 keV in the EPIC spectra, and other features in the soft X-rays. These soft X-rays features are not properly taken care by the cooling flow component, for which we modelled (model M1 and M2) them after an extra optically thin collisionally ionised plasma emission component. The temperature of this component signifies that this emission is coming from the base of the PSR.


As an alternative scenario, the photoionised warm absorber describes the soft X-ray features without the necessity of extra collisionally excited emission. Using Chandra grating data, \cite{Islam_2021} has shown the presnece of photoionised pre-shock flow for two IPs (V1223 Sgr and NY Lup). Ionisation parameter ($\xi=L/nr^{2}$) is defined as a ratio of ionisation flux to ionisation density where emission with flux $L/r^{2}$ (modulo 4$\pi$) ionises the cloud of density $n$ at distance r. This parameter is independent of the specific mass accretion rate. The 4$\pi$ factor denotes the geometrical effect for an extended cloud covering an extended source. Geometrical effects can reduce the value of the ionisation parameter significantly. Based on the photoionised warm absorber model developed by \cite{Islam_2021}, we modelled our spectra using model M3, and found a similarly good fit in comparison with model M1 and M2. The phase resolved spectral fitting using model M3 gives consistent values of ionisation parameter for phases (a), (c), and (d) within statistical uncertainty. The low value for this parameter ($\sim-1.3$) during the phase of the lowest covering fraction of complex absorber, i.e (c), is because of the geometrical effects, like the orientation and fraction of the absorber available to be ionised.

At this point, within the limited resolution of EPIC spectra and poor signal to noise ratio of RGS data, the two scenarios - presence of photoionised absorption by warm pres-shock flow; and neutral absorber with extra collisionally ionised emission from bottom of PSR, seem equally plausible from statistical perspective. A high quality grating data will reveal ionised emissions as well as absorption edges, which are important for securing the presence of a complex warm absorber.

\subsection{White dwarf mass}

The upper temperature of the cooling flow model represents the shock temperature, i.e the temperature of the shock-front on the top of the PSR. For all three model variants (M1, M2 and M3), the values of this parameter are mutually consistent with each other and well-constrained. Assuming the shock temperature ($31.7_{-3.5}^{+3.3}$ keV) from model M3 (minimum value of reduced $\chi^{2}$), we estimate a 
WD mass of $M_{\text{WD}}=0.74_{-0.05}^{+0.04} M_{\odot}$ and a WD radius of $R_{\text{WD}}=7.4_{-0.3}^{+0.4}\times10^8\;cm$. We have used WD mass-radius relationship from \cite{Nauenberg_1972}, and the formula for shock temperature with WD mass and radius (for eg. \cite{Mukai_2017}). Since, this value is measured from the shock temperature, obtained using broadband spectra, it is expected to give a fine estimate of the WD mass. However, for a negligible value of reflection amplitude, indicating a tall shock height, this mass can be somewhat underestimated \citep{Oliveira_2019}.  

The unabsorbed flux in 0.3-40.0 keV range is $F=2.11_{-0.07}^{+0.08}\times10^{-11}\;erg\;cm^{-2}\;s^{-1}$, of which $\sim33\%$ is contributed from 10-40 keV band. The corresponding luminosity (0.3-40.0 keV) is $\sim8.44\times10^{32}\;erg s^{-1}$, where distance to the source, $d=578pc$ \citep{Gaia2}.

Accretion luminosity can be represented in terms of the mass accretion rate ($\dot{M}$) as \citep{Frank_2002}
\begin{center}
    
$L_{acce}=\dfrac{GM_{\text{WD}}\dot{M}}{R_{\text{WD}}}$

\end{center}

Using our obtained values of luminosity, mass, and radius of the WD, we calculate $\dot{M} \sim 6.33\times10^{15}\;g s^{-1} \sim 9.98\times10^{-11} M_{\odot} yr^{-1}$. This value is in line with the $M_{\odot}$ seen for other asynchronous polars; for eg. - V1432 Aql: $\sim3\times10^{15}\;g s^{-1}$ \citep{Rana_2005}; CD Ind: $\sim8\times10^{14}\;g s^{-1}$ \citep{Dutta_2022}; V1500 Cyg: $\sim6\times10^{14}\;g s^{-1}$ \citep{Harrison_2016}, and BY Cam: $\sim6\times10^{16}\;g s^{-1}$ \citep{Done_1998}.

\subsection{Possibility of formation of Disc}

Past studies of Paloma \citep{Schwarz_2007, Littlefield_2022} have favoured a disc-less scenario mainly based on the presence of $2\omega-\Omega$ peak \citep{Wynn_1992} in their periodgram obtained from photometric observation. We also notice the presence of $2\omega-\Omega$ peak in the FPMA Lomb-Scargle power spectra (Fig \ref{Fig:fpmapowerspec}), but that peak is much weaker compared to $\Omega$ peak and also suppressed in CELANed power spectra. However, the power spectra presented in Figures 2 and 3 show presence of  $\Omega+2\omega$ peak which appears with relatively higher power in power spectra of both the telescopes, and in Lomb-Scargle as well as CLEANed algorithm. This frequency peak is among the set of frequency peaks, which can be produced in a purely disc-fed system with asymmetric dipole condition and additional orbital modulations (like stream-disc impact) scenario \citep{Norton_1996}. There are other expected frequency peaks for this scenario, for eg. $\omega+\Omega$ is noticed in the FPMA power spectra (but unresolved in PN power spectra). Surprisingly, the expected lower orbital sideband counterpart $\omega-\Omega$ is not noticed in the FPMA power spectra (PN power spectra is anyway unable to detect it due to smaller duration of lightcurve), and the spin harmonic $2\omega$ and its lower orbital sideband $2\omega-\Omega$ appear with small power in the FPMA Lomb-Scargle power spectra. Therefore, it is difficult to conclusively argue for disc-fed or disc-less system, as evidenced from our power spectra. Another alternate argument for the formation of disc could be made by comparing the magnetospheric radius with the other binary parameters (circularisation radius, distance of closest approach of free fall stream). Unfortunately, we could not do so because of the unavailability of the measured magnetic fields of the WD in Paloma, as of now.

\section{Summary and Conclusions}
\label{Sec:conclusion}

The X-ray broadband data of Paloma reveals interesting properties of source, some of which opens up new possibilities related to the accretion mechanism for this source. In the light of our current data, we summarize our findings in this section.

\begin{itemize}
    \item Emission from at least one WD pole is visible at any point during the orbital cycle. The complex intrinsic absorbers play crucial role in the system, which is most likely be contributed from accretion curtain/accretion stream and absorbing material due to stream-disc/stream-magnetospheric interaction.
    \item The contribution of the complex intrinsic absorber changes significantly over a complete orbital cycle with maximum contribution during orbital phase $\phi=0.1-0.22$ and minimum during phase $\phi=0.5-0.8$.
    \item Power spectra show that the orbital frequency peak is much stronger than the spin frequency peak. Strong orbital modulation possibly resulted from presence of dense blob fixed in orbital frame. Literature on this source showed spin modulations change significantly over beat cycle, which is the reason behind weak power at spin frequency in the power spectra.
    \item From the power spectra of the source, it is difficult to conclusively argue for presence or absence of partial accretion disk.
    \item The WD mass is estimated to be $0.74_{-0.05}^{+0.04} M_{\odot}$ using the shock temperature, measured directly from the broadband spectra using multitemperature emission model for the PSR.
    \item The features in the soft X-rays can be explained with similar statistical goodness either by the extra optically thin collisionally excited cold plasma emission from PSR, or by using the photoionised absorption from the warm pre-shock flow. To definitively favor one possibility, good quality high resolution data are required in the soft to medium X-rays.
    \item O-VII K$_{\alpha}$ emission line is noticed in soft X-rays. The neutral Fe K$_{\alpha}$ is weaker compared to ionised lines, and presumably the neutral line strength is mostly contributed from strong intrinsic absorber. We also measured reflection amplitude but with a poor constraint, and a small statistical improvement to the broadband fitting,  suggesting presence of weak reflection in the system.
\end{itemize}

\section{Acknowledgments}

We would like to thank the referee for valuable comments that further improved the manuscript. This research has made use of the data obtained from NuSTAR telescope, operated jointly by CalTech and NASA, and XMM-Newton telescope, operated by ESA. We thank the NuSTAR science operation team and XMM-Newton science operation team for providing the data. We acknowledge the members at the helpdesk, maintained by High Energy Astrophysics Science Archive Research Center (HEASARC) and the XMM-Newton helpdesk team members for providing necessary support. AD thanks Nazma Islam for providing useful infomation regarding \texttt{zxipab}. AD also mentions other helps provided by T.G., H.M., and K.R. during the work. 

The data used for analysis in this article are publicly available in NASA's High Energy Astrophysics Science Archive Research Center (HEASARC) archive (\url{https://heasarc.gsfc.nasa.gov/docs/archive.html}) and XMM-Newton Science archive (\url{http://nxsa.esac.esa.int/nxsa-web/#search}).

\facilities{XMM-Newton, NuSTAR}



\software{HEASoft (\cite{Heasarc_2014}, \url{https://heasarc.gsfc.nasa.gov/docs/software/heasoft/}),  
          XMM-Newton SAS (\url{https://www.cosmos.esa.int/web/xmm-newton/sas}), 
          STARLINK (\citet{Currie_2014} \url{https://starlink.eao.hawaii.edu/starlink})
          }

\bibliography{sample631}{}

\begin{thebibliography}{}
\expandafter\ifx\csname natexlab\endcsname\relax\def\natexlab#1{#1}\fi
\providecommand{\url}[1]{\href{#1}{#1}}
\providecommand{\dodoi}[1]{doi:~\href{http://doi.org/#1}{\nolinkurl{#1}}}
\providecommand{\doeprint}[1]{\href{http://ascl.net/#1}{\nolinkurl{http://ascl.net/#1}}}
\providecommand{\doarXiv}[1]{\href{https://arxiv.org/abs/#1}{\nolinkurl{https://arxiv.org/abs/#1}}}

\bibitem[{{Arnaud}(1996)}]{Arnaud_1996}
{Arnaud}, K.~A. 1996, in Astronomical Society of the Pacific Conference Series,
  Vol. 101, Astronomical Data Analysis Software and Systems V, ed. G.~H.
  {Jacoby} \& J.~{Barnes}, 17

\bibitem[{{Currie} {et~al.}(2014){Currie}, {Berry}, {Jenness}, {Gibb}, {Bell},
  \& {Draper}}]{Currie_2014}
{Currie}, M.~J., {Berry}, D.~S., {Jenness}, T., {et~al.} 2014, in Astronomical
  Society of the Pacific Conference Series, Vol. 485, Astronomical Data
  Analysis Software and Systems XXIII, ed. N.~{Manset} \& P.~{Forshay}, 391

\bibitem[{{Done} \& {Magdziarz}(1998)}]{Done_1998}
{Done}, C., \& {Magdziarz}, P. 1998, \mnras, 298, 737,
  \dodoi{10.1046/j.1365-8711.1998.01636.x}

\bibitem[{{Dutta} \& {Rana}(2022)}]{Dutta_2022}
{Dutta}, A., \& {Rana}, V. 2022, \mnras, 511, 4981,
  \dodoi{10.1093/mnras/stac296}

\bibitem[{{Ezuka} \& {Ishida}(1999)}]{Ishida_1999}
{Ezuka}, H., \& {Ishida}, M. 1999, \apjs, 120, 277, \dodoi{10.1086/313181}

\bibitem[{{Frank} {et~al.}(2002){Frank}, {King}, \& {Raine}}]{Frank_2002}
{Frank}, J., {King}, A., \& {Raine}, D.~J. 2002, {Accretion Power in
  Astrophysics: Third Edition}

\bibitem[{{Gabriel} {et~al.}(2004){Gabriel}, {Denby}, {Fyfe}, {Hoar}, {Ibarra},
  {Ojero}, {Osborne}, {Saxton}, {Lammers}, \& {Vacanti}}]{Gabriel_2004}
{Gabriel}, C., {Denby}, M., {Fyfe}, D.~J., {et~al.} 2004, in Astronomical
  Society of the Pacific Conference Series, Vol. 314, Astronomical Data
  Analysis Software and Systems (ADASS) XIII, ed. F.~{Ochsenbein}, M.~G.
  {Allen}, \& D.~{Egret}, 759

\bibitem[{{Gaia Collaboration} {et~al.}(2018){Gaia Collaboration}, {Brown},
  {Vallenari}, {Prusti}, {de Bruijne}, {Babusiaux}, {Bailer-Jones}, {Biermann},
  {Evans}, {Eyer}, {Jansen}, {Jordi}, {Klioner}, {Lammers}, {Lindegren},
  {Luri}, {Mignard}, {Panem}, {Pourbaix}, {Randich}, {Sartoretti}, {Siddiqui},
  {Soubiran}, {van Leeuwen}, {Walton}, {Arenou}, {Bastian}, {Cropper},
  {Drimmel}, {Katz}, {Lattanzi}, {Bakker}, {Cacciari}, {Casta{\~n}eda},
  {Chaoul}, {Cheek}, {De Angeli}, {Fabricius}, {Guerra}, {Holl}, {Masana},
  {Messineo}, {Mowlavi}, {Nienartowicz}, {Panuzzo}, {Portell}, {Riello},
  {Seabroke}, {Tanga}, {Th{\'e}venin}, {Gracia-Abril}, {Comoretto},
  {Garcia-Reinaldos}, {Teyssier}, {Altmann}, {Andrae}, {Audard},
  {Bellas-Velidis}, {Benson}, {Berthier}, {Blomme}, {Burgess}, {Busso},
  {Carry}, {Cellino}, {Clementini}, {Clotet}, {Creevey}, {Davidson}, {De
  Ridder}, {Delchambre}, {Dell'Oro}, {Ducourant},
  {Fern{\'a}ndez-Hern{\'a}ndez}, {Fouesneau}, {Fr{\'e}mat}, {Galluccio},
  {Garc{\'\i}a-Torres}, {Gonz{\'a}lez-N{\'u}{\~n}ez}, {Gonz{\'a}lez-Vidal},
  {Gosset}, {Guy}, {Halbwachs}, {Hambly}, {Harrison}, {Hern{\'a}ndez},
  {Hestroffer}, {Hodgkin}, {Hutton}, {Jasniewicz}, {Jean-Antoine-Piccolo},
  {Jordan}, {Korn}, {Krone-Martins}, {Lanzafame}, {Lebzelter}, {L{\"o}ffler},
  {Manteiga}, {Marrese}, {Mart{\'\i}n-Fleitas}, {Moitinho}, {Mora}, {Muinonen},
  {Osinde}, {Pancino}, {Pauwels}, {Petit}, {Recio-Blanco}, {Richards},
  {Rimoldini}, {Robin}, {Sarro}, {Siopis}, {Smith}, {Sozzetti}, {S{\"u}veges},
  {Torra}, {van Reeven}, {Abbas}, {Abreu Aramburu}, {Accart}, {Aerts},
  {Altavilla}, {{\'A}lvarez}, {Alvarez}, {Alves}, {Anderson}, {Andrei},
  {Anglada Varela}, {Antiche}, {Antoja}, {Arcay}, {Astraatmadja}, {Bach},
  {Baker}, {Balaguer-N{\'u}{\~n}ez}, {Balm}, {Barache}, {Barata}, {Barbato},
  {Barblan}, {Barklem}, {Barrado}, {Barros}, {Barstow}, {Bartholom{\'e}
  Mu{\~n}oz}, {Bassilana}, {Becciani}, {Bellazzini}, {Berihuete}, {Bertone},
  {Bianchi}, {Bienaym{\'e}}, {Blanco-Cuaresma}, {Boch}, {Boeche}, {Bombrun},
  {Borrachero}, {Bossini}, {Bouquillon}, {Bourda}, {Bragaglia}, {Bramante},
  {Breddels}, {Bressan}, {Brouillet}, {Br{\"u}semeister}, {Brugaletta},
  {Bucciarelli}, {Burlacu}, {Busonero}, {Butkevich}, {Buzzi}, {Caffau},
  {Cancelliere}, {Cannizzaro}, {Cantat-Gaudin}, {Carballo}, {Carlucci},
  {Carrasco}, {Casamiquela}, {Castellani}, {Castro-Ginard}, {Charlot},
  {Chemin}, {Chiavassa}, {Cocozza}, {Costigan}, {Cowell}, {Crifo}, {Crosta},
  {Crowley}, {Cuypers}, {Dafonte}, {Damerdji}, {Dapergolas}, {David}, {David},
  {de Laverny}, {De Luise}, {De March}, {de Martino}, {de Souza}, {de Torres},
  {Debosscher}, {del Pozo}, {Delbo}, {Delgado}, {Delgado}, {Di Matteo},
  {Diakite}, {Diener}, {Distefano}, {Dolding}, {Drazinos}, {Dur{\'a}n},
  {Edvardsson}, {Enke}, {Eriksson}, {Esquej}, {Eynard Bontemps}, {Fabre},
  {Fabrizio}, {Faigler}, {Falc{\~a}o}, {Farr{\`a}s Casas}, {Federici},
  {Fedorets}, {Fernique}, {Figueras}, {Filippi}, {Findeisen}, {Fonti},
  {Fraile}, {Fraser}, {Fr{\'e}zouls}, {Gai}, {Galleti}, {Garabato},
  {Garc{\'\i}a-Sedano}, {Garofalo}, {Garralda}, {Gavel}, {Gavras}, {Gerssen},
  {Geyer}, {Giacobbe}, {Gilmore}, {Girona}, {Giuffrida}, {Glass}, {Gomes},
  {Granvik}, {Gueguen}, {Guerrier}, {Guiraud}, {Guti{\'e}rrez-S{\'a}nchez},
  {Haigron}, {Hatzidimitriou}, {Hauser}, {Haywood}, {Heiter}, {Helmi}, {Heu},
  {Hilger}, {Hobbs}, {Hofmann}, {Holland}, {Huckle}, {Hypki}, {Icardi},
  {Jan{\ss}en}, {Jevardat de Fombelle}, {Jonker}, {Juh{\'a}sz}, {Julbe},
  {Karampelas}, {Kewley}, {Klar}, {Kochoska}, {Kohley}, {Kolenberg},
  {Kontizas}, {Kontizas}, {Koposov}, {Kordopatis}, {Kostrzewa-Rutkowska},
  {Koubsky}, {Lambert}, {Lanza}, {Lasne}, {Lavigne}, {Le Fustec}, {Le
  Poncin-Lafitte}, {Lebreton}, {Leccia}, {Leclerc}, {Lecoeur-Taibi},
  {Lenhardt}, {Leroux}, {Liao}, {Licata}, {Lindstr{\o}m}, {Lister}, {Livanou},
  {Lobel}, {L{\'o}pez}, {Managau}, {Mann}, {Mantelet}, {Marchal}, {Marchant},
  {Marconi}, {Marinoni}, {Marschalk{\'o}}, {Marshall}, {Martino}, {Marton},
  {Mary}, {Massari}, {Matijevi{\v{c}}}, {Mazeh}, {McMillan}, {Messina},
  {Michalik}, {Millar}, {Molina}, {Molinaro}, {Moln{\'a}r}, {Montegriffo},
  {Mor}, {Morbidelli}, {Morel}, {Morris}, {Mulone}, {Muraveva}, {Musella},
  {Nelemans}, {Nicastro}, {Noval}, {O'Mullane}, {Ord{\'e}novic},
  {Ord{\'o}{\~n}ez-Blanco}, {Osborne}, {Pagani}, {Pagano}, {Pailler},
  {Palacin}, {Palaversa}, {Panahi}, {Pawlak}, {Piersimoni}, {Pineau}, {Plachy},
  {Plum}, {Poggio}, {Poujoulet}, {Pr{\v{s}}a}, {Pulone}, {Racero}, {Ragaini},
  {Rambaux}, {Ramos-Lerate}, {Regibo}, {Reyl{\'e}}, {Riclet}, {Ripepi}, {Riva},
  {Rivard}, {Rixon}, {Roegiers}, {Roelens}, {Romero-G{\'o}mez}, {Rowell},
  {Royer}, {Ruiz-Dern}, {Sadowski}, {Sagrist{\`a} Sell{\'e}s}, {Sahlmann},
  {Salgado}, {Salguero}, {Sanna}, {Santana-Ros}, {Sarasso}, {Savietto},
  {Schultheis}, {Sciacca}, {Segol}, {Segovia}, {S{\'e}gransan}, {Shih},
  {Siltala}, {Silva}, {Smart}, {Smith}, {Solano}, {Solitro}, {Sordo}, {Soria
  Nieto}, {Souchay}, {Spagna}, {Spoto}, {Stampa}, {Steele},
  {Steidelm{\"u}ller}, {Stephenson}, {Stoev}, {Suess}, {Surdej}, {Szabados},
  {Szegedi-Elek}, {Tapiador}, {Taris}, {Tauran}, {Taylor}, {Teixeira},
  {Terrett}, {Teyssandier}, {Thuillot}, {Titarenko}, {Torra Clotet}, {Turon},
  {Ulla}, {Utrilla}, {Uzzi}, {Vaillant}, {Valentini}, {Valette}, {van Elteren},
  {Van Hemelryck}, {van Leeuwen}, {Vaschetto}, {Vecchiato}, {Veljanoski},
  {Viala}, {Vicente}, {Vogt}, {von Essen}, {Voss}, {Votruba}, {Voutsinas},
  {Walmsley}, {Weiler}, {Wertz}, {Wevers}, {Wyrzykowski}, {Yoldas},
  {{\v{Z}}erjal}, {Ziaeepour}, {Zorec}, {Zschocke}, {Zucker}, {Zurbach}, \&
  {Zwitter}}]{Gaia2}
{Gaia Collaboration}, {Brown}, A.~G.~A., {Vallenari}, A., {et~al.} 2018, \aap,
  616, A1, \dodoi{10.1051/0004-6361/201833051}

\bibitem[{{Harrison} {et~al.}(2013){Harrison}, {Craig}, {Christensen},
  {Hailey}, {Zhang}, {Boggs}, {Stern}, {Cook}, {Forster}, {Giommi},
  {Grefenstette}, {Kim}, {Kitaguchi}, {Koglin}, {Madsen}, {Mao}, {Miyasaka},
  {Mori}, {Perri}, {Pivovaroff}, {Puccetti}, {Rana}, {Westergaard}, {Willis},
  {Zoglauer}, {An}, {Bachetti}, {Barri{\`e}re}, {Bellm}, {Bhalerao},
  {Brejnholt}, {Fuerst}, {Liebe}, {Markwardt}, {Nynka}, {Vogel}, {Walton},
  {Wik}, {Alexander}, {Cominsky}, {Hornschemeier}, {Hornstrup}, {Kaspi},
  {Madejski}, {Matt}, {Molendi}, {Smith}, {Tomsick}, {Ajello}, {Ballantyne},
  {Balokovi{\'c}}, {Barret}, {Bauer}, {Blandford}, {Brandt}, {Brenneman},
  {Chiang}, {Chakrabarty}, {Chenevez}, {Comastri}, {Dufour}, {Elvis}, {Fabian},
  {Farrah}, {Fryer}, {Gotthelf}, {Grindlay}, {Helfand}, {Krivonos}, {Meier},
  {Miller}, {Natalucci}, {Ogle}, {Ofek}, {Ptak}, {Reynolds}, {Rigby},
  {Tagliaferri}, {Thorsett}, {Treister}, \& {Urry}}]{Harrison_2013}
{Harrison}, F.~A., {Craig}, W.~W., {Christensen}, F.~E., {et~al.} 2013, \apj,
  770, 103, \dodoi{10.1088/0004-637X/770/2/103}

\bibitem[{{Harrison} \& {Campbell}(2016)}]{Harrison_2016}
{Harrison}, T.~E., \& {Campbell}, R.~K. 2016, \mnras, 459, 4161,
  \dodoi{10.1093/mnras/stw961}

\bibitem[{{HI4PI Collaboration} {et~al.}(2016){HI4PI Collaboration}, {Ben
  Bekhti}, {Fl{\"o}er}, {Keller}, {Kerp}, {Lenz}, {Winkel}, {Bailin},
  {Calabretta}, {Dedes}, {Ford}, {Gibson}, {Haud}, {Janowiecki}, {Kalberla},
  {Lockman}, {McClure-Griffiths}, {Murphy}, {Nakanishi}, {Pisano}, \&
  {Staveley-Smith}}]{HI4PI_2016}
{HI4PI Collaboration}, {Ben Bekhti}, N., {Fl{\"o}er}, L., {et~al.} 2016, \aap,
  594, A116, \dodoi{10.1051/0004-6361/201629178}

\bibitem[{{Islam} \& {Mukai}(2021)}]{Islam_2021}
{Islam}, N., \& {Mukai}, K. 2021, \apj, 919, 90,
  \dodoi{10.3847/1538-4357/ac134e}

\bibitem[{{Jansen} {et~al.}(2001){Jansen}, {Lumb}, {Altieri}, {Clavel}, {Ehle},
  {Erd}, {Gabriel}, {Guainazzi}, {Gondoin}, {Much}, {Munoz}, {Santos},
  {Schartel}, {Texier}, \& {Vacanti}}]{Jansen_2001}
{Jansen}, F., {Lumb}, D., {Altieri}, B., {et~al.} 2001, \aap, 365, L1,
  \dodoi{10.1051/0004-6361:20000036}

\bibitem[{{Joshi} {et~al.}(2016){Joshi}, {Pandey}, {Singh}, \&
  {Agrawal}}]{Joshi_2016}
{Joshi}, A., {Pandey}, J.~C., {Singh}, K.~P., \& {Agrawal}, P.~C. 2016, \apj,
  830, 56, \dodoi{10.3847/0004-637X/830/2/56}

\bibitem[{{Littlefield} {et~al.}(2019){Littlefield}, {Garnavich}, {Mukai},
  {Mason}, {Szkody}, {Kennedy}, {Myers}, \& {Schwarz}}]{Littlefield_2019}
{Littlefield}, C., {Garnavich}, P., {Mukai}, K., {et~al.} 2019, \apj, 881, 141,
  \dodoi{10.3847/1538-4357/ab2a17}

\bibitem[{{Littlefield} {et~al.}(2022){Littlefield}, {Hoard}, {Garnavich},
  {Szkody}, {Mason}, {Scaringi}, {Ilkiewicz}, {Kennedy}, {Rappaport}, \&
  {Jayaraman}}]{Littlefield_2022}
{Littlefield}, C., {Hoard}, D.~W., {Garnavich}, P., {et~al.} 2022, arXiv
  e-prints, arXiv:2205.02863.
\newblock \doarXiv{2205.02863}

\bibitem[{{Lomb}(1976)}]{Lomb_1976}
{Lomb}, N.~R. 1976, \apss, 39, 447, \dodoi{10.1007/BF00648343}

\bibitem[{{Lopes de Oliveira} \& {Mukai}(2019)}]{Oliveira_2019}
{Lopes de Oliveira}, R., \& {Mukai}, K. 2019, \apj, 880, 128,
  \dodoi{10.3847/1538-4357/ab2b41}

\bibitem[{{Mukai}(2017)}]{Mukai_2017}
{Mukai}, K. 2017, \pasp, 129, 062001, \dodoi{10.1088/1538-3873/aa6736}

\bibitem[{{Mukai} {et~al.}(2003){Mukai}, {Kinkhabwala}, {Peterson}, {Kahn}, \&
  {Paerels}}]{Mukai_2003}
{Mukai}, K., {Kinkhabwala}, A., {Peterson}, J.~R., {Kahn}, S.~M., \& {Paerels},
  F. 2003, \apjl, 586, L77, \dodoi{10.1086/374583}

\bibitem[{{Mukai} {et~al.}(2015){Mukai}, {Rana}, {Bernardini}, \& {de
  Martino}}]{Mukai_2015}
{Mukai}, K., {Rana}, V., {Bernardini}, F., \& {de Martino}, D. 2015, \apjl,
  807, L30, \dodoi{10.1088/2041-8205/807/2/L30}

\bibitem[{{Mushotzky} \& {Szymkowiak}(1988)}]{Mushotzky_1988}
{Mushotzky}, R.~F., \& {Szymkowiak}, A.~E. 1988, in NATO Advanced Study
  Institute (ASI) Series C, Vol. 229, Cooling Flows in Clusters and Galaxies,
  ed. A.~C. {Fabian}, 53, \dodoi{10.1007/978-94-009-2953-1\_6}

\bibitem[{{Nasa High Energy Astrophysics Science Archive Research Center
  (Heasarc)}(2014)}]{Heasarc_2014}
{Nasa High Energy Astrophysics Science Archive Research Center (Heasarc)}.
  2014, {HEAsoft: Unified Release of FTOOLS and XANADU}, Astrophysics Source
  Code Library, record ascl:1408.004.
\newblock \doeprint{1408.004}

\bibitem[{{Nauenberg}(1972)}]{Nauenberg_1972}
{Nauenberg}, M. 1972, \apj, 175, 417, \dodoi{10.1086/151568}

\bibitem[{{Norton} {et~al.}(1996){Norton}, {Beardmore}, \&
  {Taylor}}]{Norton_1996}
{Norton}, A.~J., {Beardmore}, A.~P., \& {Taylor}, P. 1996, \mnras, 280, 937,
  \dodoi{10.1093/mnras/280.3.937}

\bibitem[{{Norton} {et~al.}(1997){Norton}, {Hellier}, {Beardmore}, {Wheatley},
  {Osborne}, \& {Taylor}}]{Norton_1997}
{Norton}, A.~J., {Hellier}, C., {Beardmore}, A.~P., {et~al.} 1997, \mnras, 289,
  362, \dodoi{10.1093/mnras/289.2.362}

\bibitem[{{Norton} {et~al.}(1992){Norton}, {Watson}, {King}, {Lehto}, \&
  {McHardy}}]{Norton_1992}
{Norton}, A.~J., {Watson}, M.~G., {King}, A.~R., {Lehto}, H.~J., \& {McHardy},
  I.~M. 1992, \mnras, 254, 705, \dodoi{10.1093/mnras/254.4.705}

\bibitem[{{Parker} {et~al.}(2005){Parker}, {Norton}, \& {Mukai}}]{Parker_2005}
{Parker}, T.~L., {Norton}, A.~J., \& {Mukai}, K. 2005, \aap, 439, 213,
  \dodoi{10.1051/0004-6361:20052887}

\bibitem[{{Rana} {et~al.}(2005){Rana}, {Singh}, {Barrett}, \&
  {Buckley}}]{Rana_2005}
{Rana}, V.~R., {Singh}, K.~P., {Barrett}, P.~E., \& {Buckley}, D.~A.~H. 2005,
  \apj, 625, 351, \dodoi{10.1086/429283}

\bibitem[{{Rana} {et~al.}(2004){Rana}, {Singh}, {Schlegel}, \&
  {Barrett}}]{Rana_2004}
{Rana}, V.~R., {Singh}, K.~P., {Schlegel}, E.~M., \& {Barrett}, P. 2004, \aj,
  127, 489, \dodoi{10.1086/380232}

\bibitem[{{Roberts} {et~al.}(1987){Roberts}, {Lehar}, \&
  {Dreher}}]{Roberts_1987}
{Roberts}, D.~H., {Lehar}, J., \& {Dreher}, J.~W. 1987, \aj, 93, 968,
  \dodoi{10.1086/114383}

\bibitem[{{Scargle}(1982)}]{Scargle_1982}
{Scargle}, J.~D. 1982, \apj, 263, 835, \dodoi{10.1086/160554}

\bibitem[{{Schwarz} {et~al.}(2007){Schwarz}, {Schwope}, {Staude}, {Rau},
  {Hasinger}, {Urrutia}, \& {Motch}}]{Schwarz_2007}
{Schwarz}, R., {Schwope}, A.~D., {Staude}, A., {et~al.} 2007, \aap, 473, 511,
  \dodoi{10.1051/0004-6361:20077684}

\bibitem[{{Str{\"u}der} {et~al.}(2001){Str{\"u}der}, {Briel}, {Dennerl},
  {Hartmann}, {Kendziorra}, {Meidinger}, {Pfeffermann}, {Reppin}, {Aschenbach},
  {Bornemann}, {Br{\"a}uninger}, {Burkert}, {Elender}, {Freyberg}, {Haberl},
  {Hartner}, {Heuschmann}, {Hippmann}, {Kastelic}, {Kemmer}, {Kettenring},
  {Kink}, {Krause}, {M{\"u}ller}, {Oppitz}, {Pietsch}, {Popp}, {Predehl},
  {Read}, {Stephan}, {St{\"o}tter}, {Tr{\"u}mper}, {Holl}, {Kemmer}, {Soltau},
  {St{\"o}tter}, {Weber}, {Weichert}, {von Zanthier}, {Carathanassis}, {Lutz},
  {Richter}, {Solc}, {B{\"o}ttcher}, {Kuster}, {Staubert}, {Abbey}, {Holland},
  {Turner}, {Balasini}, {Bignami}, {La Palombara}, {Villa}, {Buttler},
  {Gianini}, {Lain{\'e}}, {Lumb}, \& {Dhez}}]{Struder_2001}
{Str{\"u}der}, L., {Briel}, U., {Dennerl}, K., {et~al.} 2001, \aap, 365, L18,
  \dodoi{10.1051/0004-6361:20000066}

\bibitem[{{Thorstensen} {et~al.}(2017){Thorstensen}, {Ringwald}, {Taylor},
  {Sheets}, {Peters}, {Skinner}, {Alper}, \& {Weil}}]{Thorstensen_2017}
{Thorstensen}, J.~R., {Ringwald}, F.~A., {Taylor}, C.~J., {et~al.} 2017,
  Research Notes of the American Astronomical Society, 1, 29,
  \dodoi{10.3847/2515-5172/aa9d2a}

\bibitem[{{Turner} {et~al.}(2001){Turner}, {Abbey}, {Arnaud}, {Balasini},
  {Barbera}, {Belsole}, {Bennie}, {Bernard}, {Bignami}, {Boer}, {Briel},
  {Butler}, {Cara}, {Chabaud}, {Cole}, {Collura}, {Conte}, {Cros}, {Denby},
  {Dhez}, {Di Coco}, {Dowson}, {Ferrando}, {Ghizzardi}, {Gianotti}, {Goodall},
  {Gretton}, {Griffiths}, {Hainaut}, {Hochedez}, {Holland}, {Jourdain},
  {Kendziorra}, {Lagostina}, {Laine}, {La Palombara}, {Lortholary}, {Lumb},
  {Marty}, {Molendi}, {Pigot}, {Poindron}, {Pounds}, {Reeves}, {Reppin},
  {Rothenflug}, {Salvetat}, {Sauvageot}, {Schmitt}, {Sembay}, {Short},
  {Spragg}, {Stephen}, {Str{\"u}der}, {Tiengo}, {Trifoglio}, {Tr{\"u}mper},
  {Vercellone}, {Vigroux}, {Villa}, {Ward}, {Whitehead}, \&
  {Zonca}}]{Turner_2001}
{Turner}, M.~J.~L., {Abbey}, A., {Arnaud}, M., {et~al.} 2001, \aap, 365, L27,
  \dodoi{10.1051/0004-6361:20000087}

\bibitem[{{Verner} {et~al.}(1996){Verner}, {Ferland}, {Korista}, \&
  {Yakovlev}}]{Verner_1996}
{Verner}, D.~A., {Ferland}, G.~J., {Korista}, K.~T., \& {Yakovlev}, D.~G. 1996,
  \apj, 465, 487, \dodoi{10.1086/177435}

\bibitem[{{Wilms} {et~al.}(2000){Wilms}, {Allen}, \& {McCray}}]{Wilms_2000}
{Wilms}, J., {Allen}, A., \& {McCray}, R. 2000, \apj, 542, 914,
  \dodoi{10.1086/317016}

\bibitem[{{Wynn} \& {King}(1992)}]{Wynn_1992}
{Wynn}, G.~A., \& {King}, A.~R. 1992, \mnras, 255, 83,
  \dodoi{10.1093/mnras/255.1.83}

\end{thebibliography}
\bibliographystyle{aasjournal}



 \end{document}